\documentclass[11pt,leqno]{article}

\usepackage[english]{babel}
\usepackage[utf8x]{inputenc}
\usepackage{amsmath,amssymb,amstext,amsthm}
\usepackage{graphicx}
\usepackage[affil-it]{authblk}
\usepackage[colorinlistoftodos]{todonotes}
\usepackage[linktocpage]{hyperref}

\newtheorem{theorem}{Theorem}[section]
\newtheorem{lemma}[theorem]{Lemma}
\newtheorem{corollary}[theorem]{Corollary}
\newtheorem{definition}[theorem]{Definition}

\newtheorem{remark}[theorem]{Remark}

\newcommand{\qeds}{\qed\vspace{.2cm}}

\newcommand{\GF}{G_\FF}
\newcommand{\GFO}{G_{\FF_0}}

\newcommand{\trans}{{\sf T}}
\newcommand{\xa}{x_{A}}
\newcommand{\xb}{x_{B}}
\newcommand{\ya}{y_{A}}
\newcommand{\yb}{y_{B}}
\newcommand{\ga}{g_{A}}
\newcommand{\gb}{g_{B}}
\newcommand{\ha}{h_{A}}
\newcommand{\hb}{h_{B}}
\newcommand{\bmat}{\mathbb{M}_n(d)}

\newcommand{\mch}{\mathcal{H}}
\renewcommand{\P}{\mathcal{P}}

\newcommand{\psd}{{\mathcal{S}_+}}

\newcommand{\qarrow}{\xrightarrow{q}}

\newcommand{\C}{\ensuremath{\mathbb{C}}}
\newcommand{\F}{\ensuremath{\mathbb{F}}}
\newcommand{\FF}{\ensuremath{\mathcal{F}}}

\DeclareMathOperator{\tr}{Tr}

\DeclareMathOperator{\rel}{rel}
\DeclareMathOperator{\vect}{vec}
\DeclareMathOperator{\mat}{mat}
\DeclareMathOperator{\rk}{rk}

\def\be{\begin{equation}}
\def\ee{\end{equation}}
\usepackage{enumerate}
\newtheorem{result}{Result}

\title{Quantum and non-signalling graph isomorphisms}

\author[1]{Albert Atserias}
\author[2]{Laura Man\v{c}inska}
\author[3]{David E.~Roberson} 
\author[4]{Robert \v{S}\'{a}mal} 
\author[5]{Simone Severini} 
\author[6]{Antonios Varvitsiotis}

\affil[1]{Universitat Polit\`{e}cnica de Catalunya, Spain}
\affil[2]{University of Bristol, United Kingdom}
\affil[3]{University College London, United Kingdom}
\affil[4]{Charles University, Czech Republic}
\affil[5]{University College London, United Kingdom and Shanghai Jiao-Tong University, China}
\affil[6]{Nanyang Technological University and Centre for Quantum Technologies, Singapore}

\begin{document}

\maketitle

\begin{abstract}
We introduce  a two-player nonlocal game, called the $(G,H)$-\emph{isomorphism game}, where classical players can win with certainty if and only if the graphs $G$ and $H$ are isomorphic.  We then define the notions of quantum and non-signalling isomorphism, by considering perfect quantum and non-signalling strategies for the $(G,H)$-isomorphism game, respectively. In the quantum case, we consider both the tensor product and commuting frameworks for nonlocal games. We prove that non-signalling isomorphism coincides with the well-studied notion of fractional isomorphism, thus giving the latter an operational interpretation. Second, we show that, in the tensor product framework, quantum isomorphism is equivalent to the feasibility of two polynomial systems in non-commuting variables, obtained by relaxing the standard integer programming formulations for graph isomorphism to Hermitian variables. On the basis of this correspondence, we show that quantum isomorphic graphs are necessarily cospectral. Finally, we provide a construction for reducing linear binary constraint system games to isomorphism games. This allows us to produce quantum isomorphic graphs that are nevertheless not isomorphic. Furthermore, it allows us to show that our two notions of quantum isomorphism, from the tensor product and commuting frameworks, are in fact distinct relations, and that the latter is undecidable. Our construction is related to the FGLSS reduction from inapproximability literature, as well as the CFI construction.
\end{abstract}

\section{Introduction}

Given graphs $G$ and $H$, an \emph{isomorphism} from $G$ to $H$ is a bijection $\varphi: V(G) \to V(H)$ such that $\varphi(g) $ is adjacent to $  \varphi(g')$ if and only if $  g $ is adjacent to $  g'$. When such an isomorphism exists, we say that $G$ and $H$ are \emph{isomorphic} and write $G \cong H$. The notion of isomorphism is central to a broad area of mathematical research encompassing algebraic and structural graph theory, but also combinatorial optimization, parameterized complexity, and logic. The \emph{graph isomorphism (GI)  problem} consists of deciding whether two graphs are isomorphic. It is a question with fundamental practical interest due to the number of problems that can be reduced to it. Additionally, the GI  problem has a central role in theoretical computer science as it is one of the few naturally defined problems in NP which is not known to be polynomial-time solvable or NP-complete. While there is a deterministic quasipolynomial algorithm for the GI problem \cite{B15}, regardless of its worst case behavior, the problem can be solved with reasonable efficiency in practice (\emph{e.g.} see \cite{MP13}). In relation to the context of this paper, it is valuable to notice that the discussion around graph isomorphism has branched into the analysis of many equivalence relations that form hierarchical structures. Prominent instances are, for example, cospectrality, fractional isomorphism, \emph{etc.} \cite{B96, G13, VH03}.

We remark here that, though we will touch on algorithmic aspects of the relations we define, this is not a paper about algorithms, and we make no claims that this work is useful for developing algorithms for the graph isomorphism problem. This work is concerned with theoretical aspects of some new and old relaxations of graph isomorphism.

\paragraph{Integer programming formulations.}  As is the case for all constraint satisfaction problems, the GI problem can be formulated as an  integer programming problem.   Our next goal is to   give two of these formulations as they are relevant to this~work.   The first one is an integer  quadratic program (IQP)  and the second one an  integer linear program (ILP). 
We note that several recent developments concerning  the graph isomorphism problem are based on hierarchies of linear programming relaxations of  the ILP   formulation for the  GI problem we give below  ({\em e.g.} see \cite{atserias,GO12}).

Consider two graphs $G$ and $H$ with adjacency matrices $A_G$ and $A_H$ respectively. Recall that the \emph{adjacency matrix}, $A_G$, of a graph $G$ is a symmetric matrix whose rows and columns are indexed by $V(G)$, and such that $A_{gg'} = 1$ if $g$ is adjacent to  $g'$, and $A_{gg'} = 0$, otherwise. Throughout this work we will only consider undirected loopless simple graphs. In the IQP below, and throughout this work, we will use $\rel(g,g')$ to denote the \emph{relationship} of $g$ and $g'$, {\em i.e.}, whether they are equal, adjacent, or distinct and non-adjacent. It is easy to verify  that $G\cong H$   if and only if there exist real scalar variables $x_{gh}$ for all $g\in V(G), h\in V(H)$ such that
 following IQP is feasible:

 \be\label{IQP}\tag{IQP}
\begin{aligned}
   x_{gh}^2 & =x_{gh}, \text{ for all } g\in V(G), h\in V(H);\\
  \sum_{h' \in V(H)} x_{gh'}  &  = \sum_{g' \in V(G)} x_{g'h} =1,  \text{ for all } g \in V(G), h \in V(H);\\
  x_{gh} x_{g'h'}  & = 0,  \text{ if } \rel(g,g') \ne \rel(h,h').
\end{aligned}
\ee

The second integer programming formulation for the GI problem is based on   {\em permutation matrices}, \emph{i.e.},  square $01$-matrices with a single 1 in every row and column.
Again,  it is straightforward  to verify that  $G\cong H$   if and only if  there exists an $n\times n$  permutation matrix $P=(p_{ij})$ such that $P^\trans A_G P = A_H$, or equivalently when the following ILP is feasible:
\be\label{ILP}\tag{ILP}
\begin{aligned}
p_{ij}^2& =p_{ij}, \text{ for all } i,j\in [n];\\
\sum_{\ell = 1}^np_{\ell j}& =\sum_{k=1}^n p_{ik}=1, \text { for all } i,j\in [n];\\
A_G P & = P A_H, \text{ where } P = (p_{ij}).
\end{aligned}
\ee

By the Birkhoff-von Neumann theorem, the convex hull of the set of $n \times n$ permutation matrices is equal to the set of $n \times n$ {\em doubly stochastic matrices}, {\em i.e.,}  entrywise nonnegative matrices where the sum of the entries in each row and column is equal to $1$. This naturally suggests the following  linear relaxation of the GI  problem. We say that $G$ and $H$ are  \emph{fractional isomorphic}, and write $G \cong_f H $, if there exists a doubly stochastic matrix $D$  such that $A_G D = D A_H$. This defines an equivalence  relation on graphs that has been studied in detail and characterized in multiple ways~\cite{fraciso}.

 \paragraph{Matrix relaxations.} In this work we focus on two natural {\em matrix relaxations} of \eqref{IQP} and \eqref{ILP}. First,  we
   consider \eqref{IQP}, where we relax  the scalar variables $x_{gh}$ to $d\times d$ Hermitian indeterminates  $X_{gh}$. This leads to the following quadratic polynomial system  in Hermitian~variables:
 
 \be\label{HIQP}\tag{${\rm IQP}_d$}
\begin{aligned}
   X_{gh}^2 & =X_{gh}, \text{ for all } g\in V(G), h\in V(H);\\
  \sum_{h' \in V(H)} X_{gh'}  &  = \sum_{g' \in V(G)} =X_{g'h} =I_d,  \text{ for all } g \in V(G), h \in V(H);\\
  X_{gh} X_{g'h'}  & = 0,  \text{ if } \rel(g,g') \ne \rel(h,h').
\end{aligned}
\ee

Note that by definition,  every family of matrices  $\{X_{gh}\}_{g,h}$ which is feasible  for \eqref{HIQP} satisfies $X_{gh}^2 =X_{gh}=X^{\dagger}_{gh}$ and thus each matrix  is an  orthogonal projector. 

For the matrix relaxation of~\eqref{ILP},  we replace the permutation matrix $P$ with a block matrix $\mathcal{P} = [[P_{ij}]]$ where each block $P_{ij}$ is a $d\times d$  orthogonal projector. Thus we consider the following program in Hermitian $d\times d$ indeterminates  $P_{ij}$:

\be\label{HILP}\tag{${\rm ILP}_d$}
\begin{aligned}
P_{ij}^2& =P_{ij}, \text{ for all } i,j\in [n];\\
\sum_{\ell =1}^nP_{\ell j}& =\sum_{k=1}^n P_{ik}=I_d, \text { for all } i,j\in [n];\\
(A_G \otimes I_d)\mathcal{P} & = \mathcal{P}(A_H \otimes I_d), \text{ where } \mathcal{P} = [[P_{ij}]].
\end{aligned}
\ee 
Note that for $d=1$ the matrix $\mathcal{P}$ is exactly a permutation matrix.

As we will see in Section~\ref{sec:quantum}, the system  \eqref{HIQP} is feasible if and only if \eqref{HILP} is feasible.  Thus the feasibility of  \eqref{HIQP} (or equivalently \eqref{HILP}) corresponds to a ``robust'' relaxation of the notion of graph isomorphism which we call quantum isomorphism (see~Definition \ref{def:isomorphisms}).

Although the term ``quantum isomorphism'' might seem unmotivated at this point, as we will see in the next  section, feasibility of \eqref{HIQP} corresponds to a relaxation of graph isomorphism based on the existence of winning quantum strategies for a certain type of game. The relaxation makes use of the mathematical formalism of quantum theory and its definition requires physical resources available in quantum mechanics (see Theorem~\ref{thm:summary}).

\subsection{Nonlocal games}
A  two-party {\em nonlocal game} includes a verifier and two players, Alice and Bob, that  devise a cooperative strategy. The game is  defined in terms  of finite input sets $X_A, X_B$   and finite output sets $Y_A,$  $Y_B$   for Alice and Bob respectively, a Boolean predicate $V: X_A\times X_B\times Y_A\times Y_B\rightarrow \{0,1\}$ and a distribution $\pi$ on $X_A\times X_B$.

In the game, the verifier  samples a pair $(\xa,\xb)\in X_A\times X_B$ using the distribution $\pi$ and sends $\xa \in X_A$ to Alice and $\xb \in X_B$ to Bob. The players  respond with $\ya \in Y_A$ and $\yb \in Y_B$ respectively. We say the players {\em  win} the game  if $V(\xa, \xb,\ya,\yb) = 1$.  

The goal of Alice and Bob  is to maximize their winning probability. In the setting of nonlocal games,  the players   are allowed to agree on a strategy beforehand, but  they cannot  communicate after they  receive their questions.
The parties only play one round of this game, but we will be concerned with strategies that win with certainty, \emph{i.e.}, probability equal to 1. We will refer to such a strategy as a \emph{winning} or \emph{perfect} strategy. Lastly, note that as we only consider perfect strategies we may assume without loss of generality  that   the distribution $\pi$ has full support.

\paragraph{Strategies for nonlocal games.} A {\em deterministic classical strategy}  for a nonlocal game is one in which Alice's response is determined by her input, and similarly for Bob. In a general classical strategy, the players may use shared randomness to determine their responses. They may also use local randomness, but this can be incorporated into the shared randomness without loss of~generality.

In this paper we focus on another family of strategies where the players are allowed to use quantum resources to determine their answers. Specifically, a  {\em quantum} strategy for a nonlocal game  allows the players to determine their answers  by  performing joint  measurements on a shared quantum state. A driving force behind the emerging field of quantum computing is that quantum nonlocal effects can lead to advantages for various distributed tasks, {\em e.g.} see~\cite{bellnonlocality}. We will introduce the mathematical formalism of quantum strategies for nonlocal games in Section~\ref{subsec:qstrats}.

In Section~\ref{subsec:nonsignstrats}, we consider the family of strategies satisfying the   {\em non-signalling} constraints (see Equation~\eqref{eq:non-signalling}).   Intuitively, the non-signalling property says that  Alice's local marginal distributions   are independent  of Bob's choice of measurement and, symmetrically, Bob's local marginal distributions  are independent   of Alice's choice of measurement. Thus, Alice cannot obtain any information about Bob's input based on her input and output, and vice versa. This is the most general class of strategies we consider in this work.
 
For any of the above classes of strategies, the typical goal is to determine the maximum (or supremum) probability of winning a given nonlocal game. This is known as the classical/quantum/non-signalling \emph{value} of the game. 

\paragraph{The $(G,H)$-isomorphism game.} Given two graphs $G$ and $H$, we now define a nonlocal game which we call the $(G,H)$-\emph{isomorphism game}, with the intent of capturing and extending the notion of graph isomorphism.  The $(G,H)$-isomorphism game is played as follows: The  verifier selects  uniformly at random a pair of  vertices $\xa, \xb \in V(G) \cup V(H)$ and sends $\xa$ to Alice and  $\xb$ to Bob respectively. The players respond with vertices $\ya,\yb \in V(G) \cup V(H)$. Throughout, we assume that $V(G)$ and $V(H)$ are disjoint so that players know which graph their vertex is from. 

The first winning condition  is that each player must respond with a vertex from the graph that the vertex they received was \emph{not} from. In other words we require that:
\begin{equation}
\xa \in V(G) \Leftrightarrow \ya \in V(H) \text{ and } \xb \in V(G) \Leftrightarrow \yb \in V(H).
\label{cond1}
\end{equation}
If  condition \eqref{cond1}  is not met, the players lose. Assuming  \eqref{cond1} holds we define   $\ga$ to be the unique vertex of $G$ among $\xa$ and $\ya$, and we  define $\gb, \ha$, and $\hb$ similarly. In order to win, the  answers  of the players must also satisfy the following conditions:
\begin{equation}\label{cond2}
\rel(\ga,\gb) = \rel(\ha,\hb),
\end{equation}
In other words,  if Alice and Bob are given the same vertex, then they must respond with the same vertex. If they  receive (non-)adjacent vertices  they must return (non-)adjacent vertices. Also, assuming that Alice  receives  $g_A$ and Bob $h_B$, if Alice outputs $h_A=h_B$ then in order to win we require that   Bob returns $g_B=g_A$. Note that we do not explicitly require that $G$ and $H$ have the same number of vertices.

\subsection{Contributions}
In this work we use the $(G,H)$-{isomorphism game} in order to capture and extend the notion of graph isomorphism. First, we show that
there exists a perfect  classical strategy for the $(G,H)$-isomorphism game if and only if $G$ and $H$ are indeed isomorphic graphs. This suggests that  by considering perfect quantum and non-signalling strategies  for  the $(G,H)$-isomorphism game  we  may  define the notions of quantum and non-signalling isomorphisms of graphs. Note that we will actually consider two different classes of quantum strategies: those from the tensor product framework of joint measurements and those from the commuting operator framework. However, we will mainly focus on the former class, and when we refer to quantum strategies we will be referring to these. We will  use ``quantum commuting strategies" to refer to the latter class of strategies. The detailed formalism of quantum strategies will be given in Section~\ref{sec:preliminaries}, and quantum commuting strategies will be introduced in Section~\ref{subsec:qcisos}.

\begin{definition}\label{def:isomorphisms}
We say that two  graphs $G$ and $H$ are \emph{quantum isomorphic}/\emph{quantum commuting isomorphic}/\emph{non-signalling isomorphic}, and write $G \cong_q H$ / $G \cong_{qc} H$ / $G \cong_{ns} H$, whenever there exists a perfect quantum/quantum commuting/non-signalling strategy for the $(G,H)$-isomorphism game.
\end{definition}

This idea of associating a nonlocal game to a constraint satisfaction problem corresponding  to a certain  graph property  and  studying its quantum and non-signalling value is not new.   This  was first done for the  chromatic number of a graph  in \cite{Cameron07} and      generalized to graph homomorphisms~in~\cite{qhomos}. 

Since every classical strategy  can be trivially considered as a  quantum strategy and  any quantum strategy is necessarily non-signalling (see Equation~\eqref{eq:quantumisns}), we have that 
$$G\cong H\Rightarrow G \cong_q H \Rightarrow G \cong_{ns} H.$$
As we will see,   neither of these  implications can be reversed.

In Section \ref{sec:ns} we focus on   non-signalling isomorphism. Based on   a result of Ramana, Scheinerman, and Ullman~\cite{fraciso}  which relates fractional isomorphism to the existence  of a common equitable partition, in Theorem~\ref{thm:ns1}   we show the following:
\begin{result}\label{thm:ns}
For any  graphs $G$ and $H$  we have  that $G\cong_f H$ if and only if $G\cong_{ns} H.$
\end{result}

It is  worth noting that there is a polynomial time algorithm for determining if two graphs are fractionally isomorphic~\cite{fraciso}. Combined with Result~\ref{thm:ns}  this implies that non-signalling isomorphism is also polynomial-time decidable. Furthermore, it is known that fractional isomorphism distinguishes almost all graphs~\cite{randiso}, and so it follows that the same holds for non-signalling and quantum isomorphism, since the latter is a more restrictive relation.

In Section~\ref{sec:quantum} we focus on quantum isomorphism. We  show that perfect quantum strategies for the isomorphism game must take a special form. This allows  us to reformulate quantum isomorphism in terms of the existence of a set of projectors satisfying certain orthogonality constraints.  Based on  this we can show the  following:

\begin{result}\label{thm:summary}
For any two graphs $G,H$ the following are equivalent:
\begin{itemize}
\item[$(i)$] $G\cong_q H$;
\item [$(ii)$] The system \eqref{HIQP} is feasible for some $d\in \mathbb{N}$;
\item[$(iii)$]  The system \eqref{HILP} is feasible for some $d\in \mathbb{N}$.
\end{itemize}
\end{result}

Specifically, we prove  the equivalence $(i)\Leftrightarrow (ii)$ in Theorem~\ref{thm:qreform} and $(i)\Leftrightarrow (iii)$ in Theorem~\ref{lem:projperm}. As a consequence of Result~\ref{thm:summary} $(iii)$    it follows that  quantum isomorphic graphs must be cospectral with cospectral complements. This allows us to conclude that quantum and non-signalling isomorphism are different relations, since there are many examples of graphs that are fractionally isomorphic but not cospectral (\emph{e.g.} any pair of $r$-regular graphs is fractionally isomorphic).

Lastly, in Section~\ref{sec:separation} we consider  the question of whether  isomorphism and quantum isomorphism are different relations.  In Theorem~\ref{thm:result3} we  show that they are indeed distinct notions:

\begin{result}\label{thm:qvsclassicalseparation}
There exist graphs that are quantum isomorphic but not isomorphic.
\end{result} 

The main ingredient in the proof  of Result~\ref{thm:qvsclassicalseparation}  is  a reduction from linear binary constraint system (BCS) games, introduced by Cleve and Mittal~\cite{clevemittal}, to isomorphism games. Specifically, we show that a linear BCS game has a perfect classical (quantum) strategy if and only if a pair of graphs constructed from the  BCS are (quantum) isomorphic. Since there exist  linear BCS  games  that have perfect quantum strategies but no perfect classical strategies, this allows us to produce pairs of graphs that are quantum isomorphic but not isomorphic. The smallest example of such a pair we are able to construct uses the Mermin magic square game, which produces two graphs on~$24$ vertices each that are quantum isomorphic but not isomorphic (see Figures~\ref{fig:qiso1} and~\ref{fig:qiso2}).

The same reduction as above can be used in the quantum commuting case, and thus we obtain that a given linear BCS game has a perfect quantum commuting strategy if and only if the corresponding pair of graphs are quantum commuting isomorphic. Using this reduction and two recent results of Slofstra~\cite{slofstra16}, we are able to prove the following:
\begin{result}\label{res:qcisonotqiso}
There exist graphs that are quantum commuting isomorphic but not quantum isomorphic. Furthermore, determining if two graphs are quantum commuting isomorphic is undecidable.
\end{result} 
It will also follow from the above that quantum commuting isomorphism and non-signalling isomorphism are distinct relations, since the latter is polynomial time decidable by its equivalence with fractional isomorphism.

\section{Preliminaries}\label{sec:preliminaries}

\paragraph{Linear algebra.} The standard basis of $\C^d$ is denoted by $\{e_i: i\in [d]\}$, where  $[d]:=\{1,\dotsc,d\}$.
For a matrix $X\in \C^{d\times d}$ we denote by $X^\dagger$  its  conjugate transpose and by $X^\trans$ its transpose.
We denote    the set of $d\times d$ Hermitian operators by  $\mch^d$. Throughout this work we equip $\mch^d$ with the Hilbert-Schmidt inner product  $\langle X,Y\rangle=\tr(X^\dagger Y)$.   A matrix   $X\in \mch^d$ is called {\em positive semidefinite (psd)}  if $\psi^\dagger X\psi\ge 0$ for all $\psi\in \C^d$.  The set of $d\times d$ psd matrices  is denoted by $\mch^d_+$.  We use the fact that for two psd matrices $X,Y\in \mch^d_+$ we have that $XY=0$ if and only if $\langle X,Y\rangle=0$.

A matrix  $E$ is called an (orthogonal) {\em projector} if it satisfies $E=E^\dagger=E^2$. We  typically omit  the  term ``orthogonal" because we will often refer to two projectors $E$ and $F$ being orthogonal (to each other) whenever they satisfy $EF = 0$. We use the fact that for any  family of projectors   $\{E_i\}_i$ satisfying $\sum_i E_i=I$ we have that $E_iE_j=0$, for all $i\ne j$.

We denote by $\bmat$ the set of $n\times n$ block matrices whose  blocks are  matrices  in  $\mch^d$.  For any family of matrices $\{E_{ij}\}_{i,j=1}^n\subseteq \mch^d$ we denote by  $[[E_{ij}]]$  the element of  $\bmat$  whose $(i,j)$-block is equal to $E_{ij}$.  The $(i,j)$-block of  a matrix $\P\in \bmat$ is denoted by $\P_{i,j}$.

\paragraph{Quantum mechanics.}

In this section we briefly 
review some basic concepts from the theory of quantum information. For additional details  we refer the reader to~\cite{NC} and references~therein.

To any quantum system one can associate a Hilbert space $\mathbb{C}^d$. The \emph{state} of the system is described by a unit vector $\psi \in \mathbb{C}^d$. Note that states that can be described in this way are actually known as \emph{pure} states, and more generally the state of a quantum system is  described by a Hermitian positive semidefinite matrix with trace equal to one. However, for quantum strategies for nonlocal games it suffices to consider only pure states, so we restrict our attention to this case. 

One can obtain classical information from a quantum system by measuring it. For the purposes of this paper, the most relevant mathematical formalism of the concept of measurement is given by a Positive Operator-Valued Measure (POVM). A POVM $\mathcal{M}$ consists of a family of Hermitian psd   matrices $\{M_i \in \mch_+^d : i \in [m]\}$ such that $\sum_{i=1}^m M_i = I$, where $m$ is some integer and $I$ is the identity matrix. According to quantum mechanics, if the measurement $\mathcal{M}$ is performed on a quantum system in state $\psi \in \mathbb{C}^d$, then the probability that outcome $i$ occurs is $\psi^\dagger M_i \psi$. We say that a measurement $\mathcal{M}$ is \emph{projective} if all of the POVM elements are projectors. Note that for any set of projectors $\{M_i : i \in [m]\}$ the condition $\sum_{i=1}^m M_i = I$ implies that the $M_i$'s are mutually orthogonal. Therefore the POVM elements of any projective measurement are orthogonal to each~other.

Consider two quantum systems ${\rm S_1}$ and ${\rm S_2}$ with corresponding state spaces $\C^{d_1}$ and $ \C^{d_2}$ respectively.   The state space of the joint system $({\rm S_1}, {\rm S_2})$ is given by  the tensor product $\C^{d_1}\otimes \C^{d_2}$.  Moreover, if the system ${\rm S_1}$ is in (pure) state $\psi_1\in \C^{d_1}$ and ${\rm S_2}$ is in (pure) state $\psi_2\in \C^{d_2}$ then the {\em joint system} is in state $\psi_1\otimes \psi_2\in \C^{d_1}\otimes \C^{d_2}$. Not every state in the joint system space $\C^{d_1} \otimes \C^{d_2}$ can be written as a tensor product. States that cannot be written as a tensor product are known as \emph{entangled} states. It is the existence of entangled states  that allows for quantum advantage in nonlocal games and many other scenarios.

If  $\{M_i\in \mch_+^{d_1} : i\in [m_1]\}$ and $\{N_j\in \mch_+^{d_2} : j\in [m_2]\}$
define  measurements on the individual systems ${\rm S_1}$ and ${\rm S_2}$  then the family of operators $\{M_i\otimes N_j\in \mch_+^{d_{1} d_{2}}:i\in [m_1], j\in [m_2]\}$ defines  a {\em product measurement} on the joint system $({\rm S_1}, {\rm S_2})$.   The  probability of  getting outcome $(i,j)\in [m_1]\times [m_2]$, when measuring the quantum state $\psi$, is equal to~$\psi^*(M_i\otimes N_j)\psi$.

It is often convenient to use the fact that any quantum state $\psi\in\C^{d}\otimes\C^{d}$ admits a so-called \emph{Schmidt decomposition}: $\psi = \sum_{i=1}^d \lambda_i\, \alpha_i \otimes \beta_i$ where $\{\alpha_i : i\in[d]\}$ and $\{\beta_i : i\in[d]\}$ are orthonormal bases of $\C^d$ and $\lambda_i \ge 0$ for all $i\in [d]$. The bases $\{\alpha_i : i\in[d]\}$ and $\{\beta_i : i\in[d]\}$ are known as the \emph{Schmidt bases} of $\psi$, and the $\lambda_i$ are its {\em Schmidt coefficients}. We say that $\psi$ has {\em full Schmidt rank} if its Schmidt coefficients are all positive. Note that one can also consider a Schmidt decomposition of states in $\mathbb{C}^{d_1} \otimes \mathbb{C}^{d_2}$ where $d_1 \ne d_2$, but for us it suffices to consider~$d_1 = d_2$.

We say that a state is  \emph{maximally entangled} if all of its Schmidt coefficients are the same. The \emph{canonical maximally entangled state} in $\C^d \otimes \C^d$ is the state $\psi_d:=\frac{1}{\sqrt{d}} \sum_{i=1}^d e_i \otimes e_i$, where $e_i$ is the $i^\text{th}$ standard basis vector. We will make use of the fact that
\be\label{eq:maxent}
\psi_d^\dagger(A\otimes B)\psi_d= \frac{1}{d} \, {\rm Tr}\left(AB^\trans\right), \text{ for all } A,B\in \C^{d\times d}.
\ee

\section{Strategies for the $(G,H)$-isomorphism game}\label{sec:isogame}

In this section we introduce   three families  of strategies for the $(G,H)$-isomorphism game (classical, quantum, and non-signalling) and study how they relate  to each other.

Given a fixed strategy for the $(G,H)$-isomorphism game, we denote by $p(\ya, \yb | \xa, \xb)$ the joint conditional probability  of Alice and Bob responding with $\ya$ and $\yb$ upon receiving  inputs $\xa$ and $\xb$ respectively.  We call such a joint conditional probability distribution a \emph{correlation}. Let $S$ be a   strategy   for a  nonlocal game  and let $p_S$ be the corresponding correlation.  An easy but important observation is that $S$ is a perfect strategy  if and only if   $p_S(\ya, \yb | \xa, \xb)=0$ whenever $\xa,\xb,\ya,\yb$ do not meet the winning conditions of the game, {\em i.e.,}
 \be\label{eq:perfectsttategies}
p_S(\ya, \yb | \xa, \xb)=0,  \text{ when }  \pi(\xa,\xb)>0 \text{ and } V(\xa, \xb,\ya,\yb) = 0.
 \ee

In particular, if we specialize \eqref{eq:perfectsttategies} to   the $(G,H)$-isomorphism game we have that the correlation $p$ corresponds to a perfect strategy if and only if 
\be\label{eq:pstartegiesconditions} 
p(\ya, \yb | \xa, \xb)=0, \text{ whenever conditions }  \eqref{cond1} \text{ or } \eqref{cond2} \text{ fail}.
\ee

As a consequence we have that any winning strategy for the $(G,H)$-isomorphism game is also a winning strategy for the $(H,G)$-isomorphism game, as well as the $(\overline{G},\overline{H})$-isomorphism game.

\subsection{Classical Strategies}

In a classical strategy, Alice and Bob are allowed to make use of shared randomness to determine how they respond. Note that this does not allow them to communicate. They may also use local randomness, but this can be incorporated into the shared randomness without loss of generality. Formally, this means that the correlation associated to a classical strategy has the form $p = \sum_i \lambda_i p_i,$
where the $\lambda_i$'s encode the shared randomness and satisfy $\lambda_i > 0$, and $\sum_i \lambda_i = 1$, and the $p_i$ are correlations arising from deterministic classical strategies, \emph{i.e.}, for each $i$, $p_i(\ya, \yb | \xa, \xb) \in \{0,1\}$ for all $\xa, \xb, \ya, \yb \in V(G) \cup V(H)$. Since whether a correlation corresponds to a winning strategy is determined by its zeros, the correlation $p$ arises from a winning strategy if and only if $p_i$ is winning for all $i$. Thus we can consider the deterministic strategy corresponding to $p_1$. A deterministic classical strategy amounts to a pair of functions, $f_A, f_B : V(G) \cup V(H) \to V(G) \cup V(H)$, which map inputs to outputs for each of Alice and Bob respectively. Assuming the strategy is winning, we have that $f_A(x), f_B(x) \in V(G) \Leftrightarrow x \in V(H)$, and that $f_A(x) = f_B(x)$ for all $x \in V(G) \cup V(H)$. Since $f_A = f_B$, we will refer to both of them as simply $f$. For $g, g' \in V(G)$, the winning conditions of the $(G,H)$-isomorphism game require that $\rel(g,g') = \rel(f(g), f(g'))$. This implies that the restriction of $f$ to $V(G)$ is an isomorphism from $G$ to an induced subgraph of $H$. Similarly, the restriction of $f$ to $V(H)$ is an isomorphism of $H$ to an induced subgraph of $G$. This is only possible if $G$ and $H$ are isomorphic and the above two restrictions of $f$ are isomorphisms. Finally, for $g^* \in V(G)$, let $h^* = f(g^*)$. The case where Alice is sent $g^*$ and Bob is sent $h^*$ allows us to conclude that $\rel(g^*, f(h^*)) = \rel(f(g^*), h^*)$. Since $h^* = f(g^*)$, the relationship between these vertices is `equality', thus $g^* = f(h^*)$. In other words, the restriction $f|_{V(G)}$ is the inverse of the restriction~$f|_{V(H)}$.

The above shows that any winning deterministic strategy for the $(G,H)$-isomorphism game corresponds to Alice and Bob responding according to a fixed isomorphism between $G$ and $H$. Moreover, any classical strategy can be decomposed as a probabilistic (or convex) combination of deterministic strategies. Therefore, classical players can win the $(G,H)$-isomorphism game only if $G$ and $H$ are indeed isomorphic.

Conversely, suppose that $\varphi : V(G) \to V(H)$ is an isomorphism of graphs $G$ and $H$. It is easy to see that if both players respond with $\varphi(g)$ upon receiving $g \in V(G)$ and respond with $\varphi^{-1}(h)$ upon receiving $h \in V(H)$, then they will win the $(G,H)$-isomorphism game. So we see that there exists a winning classical strategy for the $(G,H)$-isomorphism game if and only if $G$ and $H$ are indeed isomorphic graphs.

\subsection{Quantum Strategies}\label{subsec:qstrats}

A  quantum strategy for the $(G,H)$-isomorphism game consists of a shared entangled state $\psi$, and POVMs  $\mathcal{E}_x = \{E_{xy} : y \in V(G) \cup V(H)\}$ for each $x \in V(G) \cup V(H)$ for Alice, and  $\mathcal{F}_x = \{F_{xy} : y \in V(G) \cup V(H)\}$ for each $x \in V(G) \cup V(H)$ for Bob. Upon receiving $\xa \in V(G) \cup V(H)$ Alice performs measurement $\mathcal{E}_{\xa}$ and obtains some outcome $\ya \in V(G) \cup V(H)$. Similarly,  upon receiving $\xb$ Bob measures $\mathcal{F}_{\xb}$   and obtains some $\yb$. The probability of Alice and Bob outputting vertices $\ya$ and $\yb$ upon receiving $\xa$ and $\xb$ respectively is given by 
\be\label{eq:qcorrelation}
p(\ya,\yb|\xa,\xb) = \psi^\dagger\left(E_{\xa\ya} \otimes F_{\xb\yb}\right)\psi.
\ee
Any correlation that can be realized as in \eqref{eq:qcorrelation} is known as a \emph{quantum correlation}.

Therefore, it follows by~\eqref{eq:pstartegiesconditions} that a quantum strategy as described above is a winning strategy for the $(G,H)$-isomorphism game if and only if 
\[\psi^\dagger\left(E_{\xa\ya} \otimes F_{\xb\yb}\right)\psi = 0,  \text{ whenever conditions }  \eqref{cond1} \text{ or } \eqref{cond2} \text{ fail}.\]

It is important to note that any classical correlation is also a quantum correlation. Indeed, any deterministic strategy can be produced by using measurements in which all but one of the POVM elements is the zero matrix. The remaining POVM element will be the identity and performing this measurement will always result in the outcome corresponding to the identity. Since any classical shared randomness can also be replicated by measurements on a shared state, this shows that any classical correlation can be produced by some quantum strategy.

\subsection{Non-signalling Strategies}\label{subsec:nonsignstrats}

Suppose that Alice and Bob are playing a nonlocal game with a quantum strategy as described in the previous section. If Alice is given input $\xa$, and Bob is given input $\xb$,  the probability that Alice obtains outcome $\ya$ when she performs  measurement $\mathcal{E}_{\xa}$ is given by:
\be\label{eq:quantumisns}
\sum_{\yb} p(\ya, \yb | \xa, \xb) = \sum_{\yb} \psi^\dagger \left(E_{\xa\ya} \otimes F_{\xb\yb}\right)\psi = \psi^\dagger \left(E_{\xa\ya} \otimes I\right)\psi,
\ee
and we  see that this does not depend on Bob's input $\xb$. Similarly, the probability of Bob obtaining a particular outcome will not be dependent on Alice's input. This property of quantum correlations is known as the \emph{non-signalling} property. 
Formally, a correlation $p(\ya, \yb | \xa, \xb)$ is {\em non-signalling}~if
\be\label{eq:non-signalling}
\begin{aligned}
\sum_{\yb} p(\ya, \yb | \xa, \xb) &= \sum_{\yb} p(\ya, \yb | \xa, \xb'), \text{ for all } \xa, \ya, \xb, \xb', \text{ and }  \\
\sum_{\ya} p(\ya, \yb | \xa, \xb) &= \sum_{\ya} p(\ya, \yb | \xa', \xb), \text{ for all } \xb, \yb, \xa, \xa'
\end{aligned}
\ee
In other words, a non-signalling correlation does not allow the two parties to send information between themselves. If it is the case that nothing, including information, can travel faster than the speed of light, then any correlation produced by sufficiently distant parties must be non-signalling. More specifically, if Alice and Bob are separated by a large enough distance, and they are required to respond to the verifier quickly enough, then we can be certain that their correlation is non-signalling.

As we have seen, all quantum correlations are non-signalling. However, the converse is not true. For instance, for input and output sets equal to $\{0,1\}$, the \emph{PR box}~\cite{PR} is the correlation given by:
\[p(y, y' | x, x') = 
\begin{cases}
\frac{1}{2}, & \text{if } y + y' \equiv  xx' \mod 2 \\
0, & \text{otherwise.} 
\end{cases}\]
One can check that this correlation is non-signalling, but it is well known~\cite{PR} that it cannot be implemented by any quantum strategy.

A general non-signalling correlation may not be physically realizable, so when we refer to non-signalling strategies, we can think of this as Alice and Bob each simply having some black box where  they enter their inputs into and which gives them their outputs. We only require that the resulting correlation produced by these boxes obeys the non-signalling condition.

Any correlation that is \emph{not} non-signalling allows Alice and Bob to communicate some information. However, this violates the definition of a nonlocal game, since one of the requirements is that the players are not allowed to communicate during the game. Thus, one of the reasons for considering non-signalling correlations is that they represent the largest class of admissible correlations for nonlocal games. More practically, since the non-signalling condition is linear, these correlations often provide tractable upper bounds on the power of quantum correlations. Indeed, in the next section we will see that we can completely characterize when two graphs are non-signalling isomorphic.

\section{Non-signalling Isomorphism}\label{sec:ns}

 Our goal in this section is to show Result~\ref{thm:ns}, \emph{i.e.}, that fractional isomorphism and non-signalling isomorphism are equivalent relations.

\subsection{Non-signalling isomorphism implies  fractional isomorphism}
To show that non-signalling isomorphism implies  fractional isomorphism we  show that one can use a non-signalling correlation that wins the $(G,H)$-isomorphism game to construct a doubly stochastic matrix $D$ satisfying $A_G D = D A_H$.

First, if $p$ is a winning non-signalling correlation for the $(G,H)$-isomorphism game, then we must have that $p(g,y|g',x) = 0$ whenever $g, g' \in V(G)$, and similarly when we replace  $G$ by $H$ and/or switch Alice and Bob's positions. Furthermore, for all $h \in V(H)$ we have that $p(g,g'|h,h) = 0$ if $g \ne g'$, and similarly with $H$ replaced by $G$. Therefore, we have the following observation:
\be
\label{eqn:sums1}
\sum_{h' \in V(H)} p(h', h' | g, g)= \sum_{g' \in V(G)} p(g', g' | h, h)=1, \text{ for all } g \in V(G), h\in V(H).
\ee

Our goal is to  use \eqref{eqn:sums1} to construct the desired  doubly stochastic matrix. Specifically, the above sums  will  correspond to its row and column sums. We need the following intermediate result.

\begin{lemma}\label{lem:switch}
Let $p$ be a winning non-signalling correlation  for the $(G,H)$-isomorphism game.~Then,
\[p(h,h|g,g) = p(g,h | h,g) = p(h,g|g,h) = p(g,g|h,h),\]
for all $g \in V(G)$, $h \in V(H)$.
\end{lemma}
\proof
Set  $V:= V(G) \cup V(H)$. For all $ g \in V(G)$ and $h \in V(H)$ we have that
\[p(h,h | g,g) = \sum_{y \in V} p(h,y | g,g) = \sum_{y \in V} p(h,y | g,h) = p(h,g | g,h),\]
where we use \eqref{eq:pstartegiesconditions} for the first equality, for the second equality we use that $p$ is non-signalling and for the third equality we again use \eqref{eq:pstartegiesconditions}. Similarly, we get that 
\[p(h,h | g,g) = \sum_{y \in V} p(y,h | g,g) = \sum_{y \in V} p(h,y | h,g) = p(g,h | h,g).\]
Lastly, by the  symmetry of $G$ and $H$  we also have that $p(g,g | h,h) = p(g,h | h, g)$. Putting everything together the  lemma follows.\qeds

Note that by combining Lemma~\ref{lem:switch}   with Equation~(\ref{eqn:sums1}) we get  that
\[|V(G)| = \sum_{g \in V(G), h \in V(H)} p(h,h | g,g) = \sum_{g \in V(G), h \in V(H)} p(g,g | h,h) = |V(H)|,\]
which was not obvious even for quantum strategies.

We can now show that two graphs which are non-signalling isomorphic are necessarily fractionally isomorphic.

\begin{lemma}\label{lem:ns2frac}
If $G$ and $H$ are non-signalling isomorphic, then they are fractionally isomorphic.
\end{lemma}
\proof
Set  $V:= V(G) \cup V(H)$  and let $A_G$ and $A_H$ be the adjacency matrices of $G$ and $H$ respectively. Define $D$ to be a matrix with rows indexed by $V(G)$ and columns by $V(H)$ such that $D_{gh} = p(h,h|g,g)$, for all $g\in V(G),h\in V(H)$. We  show that $D$ is doubly stochastic and  satisfies $A_G D =~DA_H$. First, the $g^\text{th}$ row sum of $D$ is given by $\sum_{h \in V(H)} p(h,h|g,g) $ which is equal to $1$
by Equation~(\ref{eqn:sums1}). Furthermore, the $h^\text{th}$ column sum of $D$ is given by:
\[\sum_{g \in V(G)} p(h,h|g,g) = \sum_{g \in V(G)} p(g,g|h,h) = 1,\]
where for the first equality we use Lemma~\ref{lem:switch} and for the second one we use \eqref{eqn:sums1}.
Since the entries of $D$ are also obviously nonnegative, we have that $D$ is doubly stochastic.

Consider the $(g,h)$-entries of the matrices $A_G D$ and $DA_H$. We have that
\begin{align*}
(A_G D)_{gh} &= \sum_{g' : g' \sim g} p(h,h|g',g') \\
(DA_H)_{gh} &= \sum_{h' : h' \sim h} p(h',h'|g,g)
\end{align*}

Making repeated use of \eqref{eq:pstartegiesconditions} and the non-signalling conditions \eqref{eq:non-signalling} we get that
\be\label{whateverrrr}
\begin{aligned}
\sum_{h' : h' \sim h} p(h',h'|g,g) &= \sum_{h' : h' \sim h} \sum_{y \in V} p(h',y|g,g) = \sum_{h' : h' \sim h} \sum_{y \in V} p(h',y|g,h) \\
&= \sum_{h' : h' \sim h} \sum_{g' : g' \sim g} p(h',g'|g,h) = \sum_{g' : g' \sim g} \sum_{h' : h' \sim h} p(h',g'|g,h) \\
&= \sum_{g' : g' \sim g} \sum_{y \in V} p(y,g'|g,h) = \sum_{g' : g' \sim g} \sum_{y \in V} p(y,g'|h,h) \\
&= \sum_{g' : g' \sim g} p(g',g'|h,h).
\end{aligned}
\ee
By Lemma~\ref{lem:switch}  we have $p(g',g'|h,h)=p(h,h|g',g')$ and thus \eqref{whateverrrr} implies that
\[\sum_{h': h' \sim h} p(h',h'|g,g) = \sum_{g': g' \sim g} p(h,h|g',g'),\]
\emph{i.e.}, that $(DA_H)_{gh} = (A_G D)_{gh}$. Therefore $A_G D = DA_H$ and $G$ and $H$ are fractionally isomorphic.\qed

\subsection{Fractional isomorphism implies  non-signalling isomorphism}

To show the converse of Lemma~\ref{lem:ns2frac}, we   use a result of Ramana, Scheinerman, and Ullman~\cite{fraciso} which shows that  fractional graph isomorphism is equivalent to deciding whether the graphs have a common equitable partition. To explain this result we first need to introduce some definitions.

Let  $\mathcal{C} = \{C_1, \ldots, C_k\}$ be  a partition of $V(G)$ for some graph $G$. The partition $\mathcal{C}$ is called \emph{equitable} if there exist numbers $c_{ij}$ for $i,j \in [k]$ such that any vertex in $C_i$ has exactly $c_{ij}$ neighbors in $C_{j}$. Note that $c_{ij}$ and $c_{ji}$ are not necessarily equal, but $c_{ij}|C_i| = c_{ji}|C_j|$. We  refer to  the numbers $c_{ij}$  as the \emph{partition numbers} of an equitable partition $\mathcal{C}$. A trivial example of this is the partition where each  part has size 1. Less trivially, if $G$ is regular,  the partition with only one cell is~equitable.

Equivalently, a partition  $\mathcal{C} = \{C_1, \ldots, C_k\}$ is equitable if for any $i \in [k]$, the subgraph induced by the vertices in $C_i$ is regular, and for any $i \ne j \in [k]$ the subgraph with vertex set $C_i \cup C_j$ and containing the edges between $C_i$ and $C_j$ is a semiregular bipartite graph.

We say that $\mathcal{C}$ and $\mathcal{D}$ have  a \emph{common equitable partition} if there exist   equitable partitions $\mathcal{C} = \{C_1, \ldots, C_k\}$ and $\mathcal{D} = \{D_1, \ldots, D_{k'}\}$ for $G$ and $H$ respectively, satisfying  $k = k'$,  $|C_i| = |D_i|$ for all $i \in [k]$, and lastly, $c_{ij} = d_{ij}$ for all $i,j \in [k]$. As an  example, if $G$ and $H$ are both $d$-regular and have the same number of vertices, then the single cell partitions form a common equitable partition, and thus any such graphs are fractionally isomorphic.

As it turns out, common equitable partitions characterize  the notion of fractional isomorphism.

\begin{theorem}[\cite{fraciso}]\label{thm:eqpart} Two graphs are fractionally isomorphic if and only if they have a common equitable partition.
\end{theorem}

We  prove the converse of Lemma~\ref{lem:ns2frac} by showing that a common equitable partition can be used to construct a non-signalling correlation that wins the $(G,H)$-isomorphism game.

\begin{lemma}\label{lem:otherdirection}
 If  $G$ and $H$ are fractionally isomorphic, then  they are non-signalling isomorphic.
\end{lemma}
\proof
As $G\cong_f H$, by Theorem~\ref{thm:eqpart} the graphs $G$ and $H$ have a common equitable partition $\mathcal{C} = \{C_1, \ldots, C_k\}$ and $\mathcal{D} = \{D_1, \ldots, D_k\}$ respectively. Let $n_i = |C_i| = |D_i|$ for all $i$ and let $c_{ij}$ for $i,j \in [k]$ be the common partition numbers. Also define $\overline{c}_{ij} := n_{j} - c_{ij} - \delta_{ij}$, where $\delta_{ij}$ is the Kronecker delta function. Note that $\overline{c}_{ij}$ is the number of non-neighbors a vertex of $C_i$ has in~$C_j$.

We  use this common equitable partition to construct a winning non-signalling correlation $p$. The idea is roughly that if Alice and Bob are given $g \in C_i$ and $g' \in C_j$ respectively such that $g\sim g'$, they should respond in a correlated manner with the endpoints of a randomly chosen edge between $D_i$ and $D_j$. Formally, for $g \in C_i$, $g' \in C_j$, $h \in D_i$, $h' \in D_j$, define a correlation $p$ as~follows:
\be\label{eq:definitionnscorr}
p(h,h'|g,g') = 
\begin{cases}
\frac{1}{n_i c_{ij}}, & \text{if } g \sim g' \ \& \ h \sim h' \\
\frac{1}{n_i \overline{c}_{ij}}, & \text{if } g \not\simeq g' \ \& \ h \not\simeq h' \\
\frac{1}{n_i}, & \text{if } g = g' \ \& \ h = h' \\
0, & \text{otherwise.}
\end{cases}
\ee
Furthermore, define  
\be\label{eq:vdvdrrbr}
p(h,h'|g,g') = p(h,g'|g,h') = p(g,h'|h,g') = p(g,g'|h,h'),\ \forall g,g',h,h',
\ee
and lastly set   $p(y,y'|x,x')$  equal to zero for all values of $x,x',y,y'$ not yet accounted for. 

It is easy to verify  that this correlation evaluates to  zero when Alice and Bob's inputs and outputs do not meet the winning conditions of the $(G,H)$-isomorphism game and thus it corresponds to a perfect strategy.  Thus, it only remains to show that $p$ is a valid non-signalling correlation, {\em i.e.,} it satisfies Equation~\eqref{eq:non-signalling}. In fact, we show that for all $\xa, \xb, \ya \in V(G) \cup V(H)$ we have:
\begin{equation}\label{eqn:marginals}
\sum_{\yb\in V} p(\ya,\yb|\xa,\xb) = 
\begin{cases}
\frac{1}{n_i}, & \text{if } (\xa \in C_i \ \& \ \ya \in D_i) \text{ or } (\xa \in D_i \ \& \ \ya \in C_i); \\
0, & \text{otherwise.}
\end{cases}
\end{equation}
and similarly when  Alice and Bob exchanged.  As this does not depend on the choice of $\xb$ it follows by definition  that the correlation $p$ is non-signalling. 

Now we proceed to prove \eqref{eqn:marginals}. If $ \xa  \in C_i \ \& \ \ya \not \in D_i$ (or  $ \xa  \in D_i \ \& \ \ya  \not\in C_i$) we have by  definition  that $\sum_{\yb} p(\ya,\yb|\xa,\xb)=~0$.
It remains to  consider   the case  $g:=\xa \in C_i $ and $ h:=\ya \in D_i$ (the case $\xa \in D_i \ \& \ \ya \in C_i$ follows similarly). For clarity of exposition we  divide  the proof  in two subcases.

\medskip 

\noindent {\bf Case 1:} If  $g':=\xb\in C_j$  it follows by \eqref{eq:definitionnscorr} that 
\be
\sum_{\yb\in V} p(h,\yb|g,g') =\sum_{h'\in D_j} p(h,h'|g,g')=
\begin{cases}
 p(h,h|g,g'), & \text{ if } g=g';\\
\sum\limits_{h'\in D_j \cap N(h) } p(h,h'|g,g'), &  \text{ if } g \sim g';\\
\sum\limits_{h'\in D_j\cap N[h]^c} p(h,h'|g,g'), & \text{ if } g \not \simeq g',\\
\end{cases}
\ee
and again by Equation \eqref{eq:definitionnscorr} this evaluates to $1/n_i$ in all  three cases.

\medskip 
\noindent {\bf Case 2:} If  $h':=\xb\in D_j$  we have that:
\be
\sum_{\yb\in V} p(h,\yb|g,h') =\sum_{g'\in C_j} p(h,g'|g,h')=\sum_{g'\in C_j} p(g,g'|h,h')=1/n_i,
\ee
where the first equality follows from \eqref{eq:definitionnscorr}, the second one from~\eqref{eq:vdvdrrbr} and the third one   by Case~1.

Lastly, we show that $p$ is a valid probability distribution. For this, let $\xa,\xb\in V$ and assume that   $\xa \in C_i$ (the case $\xa\in D_i$ is similar).~Then, we have that
\[\sum_{\ya\in V}\sum_{\yb\in V} p(\ya,\yb | \xa,\xb)= \sum_{\ya\in D_i}\sum_{\yb\in V} p(\ya,\yb | \xa,\xb)=\sum_{\ya\in D_i}\frac{1}{n_i}=1,\]
where for the second equality we used \eqref{eqn:marginals}.\qeds

Combining Lemma~\ref{lem:ns2frac} with Lemma~\ref{lem:otherdirection} we immediately get   Result~\ref{thm:ns}.

\begin{theorem}\label{thm:ns1}
For any  graphs $G$ and $H$  we have  that $G\cong_f H$ if and only if $G\cong_{ns} H.$
\end{theorem}

As mentioned above, a common example of graphs that are fractionally isomorphic but not isomorphic is any pair of non-isomorphic $d$-regular graphs on $n$ vertices. This makes it seem like fractional isomorphism is a quite coarse relaxation of isomorphism. But in fact it is known~\cite{randiso} that asymptotically almost surely every graph is not fractionally isomorphic to any graph that it is not also isomorphic to. Since non-signalling/fractional isomorphism is the coarsest relation we will consider in this work, the same holds for all the other relations we will see. 

It is worth noting that this is quite different from the related graph-based nonlocal game known as the $(G,H)$-homomorphism game~\cite{qhomos, RobersonThesis}. As its name suggests, this game can be won classically if and only if there exists a homomorphism (adjacency-preserving map) from $G$ to $H$. However, if non-signalling strategies are allowed, the game becomes trivial: it can always be won as long as $H$ has at least one edge~\cite{qhomos}.

\section{Quantum graph isomorphism}\label{sec:quantum}
In this section we prove Result~\ref{thm:summary}, {\em i.e.,} we show that quantum isomorphism coincides with existence of feasible solutions to the programs  \eqref{HIQP} and \eqref{HILP}.

\subsection{Common techniques for analyzing quantum strategies}  

Recall that  a correlation $p(y,y'|x,x') $  is  quantum if it can be generated by a quantum strategy, {\em i.e.,}   if there exists a quantum state $\psi\in \C^d\otimes \C^d$ and  measurements   $\{E_{xy}\}_y$  and $\{F_{xy}\}_y$ such that
\be\label{eq:qcorrelationform}
p(y,y'|x,x')=\psi^\dagger\left(E_{xy} \otimes F_{x'y'}\right)\psi,\  \forall x,x',y,y'.
\ee
It is well-known that we may assume without loss of generality that the state  $\psi$ has full Schmidt rank.  To see this let    $\psi = \sum_{i=1}^{d'} \lambda_i\, \alpha_i \otimes \beta_i$ be the  Schmidt decomposition for $\psi$, where $\{\alpha_i\}_{i=1}^{d'}$ and $\{\beta_i\}_{i=1}^{d'}$ are orthonormal bases of $\C^d$ and $\lambda_i >0$ for all $i\in [d']$.  Consider the isometries $U_A=\sum_{i=1}^{d'}e_ia_i^\dagger$ and $U_B=\sum_{i=1}^{d'}e_ib_i^\dagger$, where $e_i \in \C^{d'}$, and define 
\be\label{eq:localisometry}
\tilde{\psi}=(U_A\otimes U_B)\psi, \quad \tilde{E}_{xy}=U_AE_{xy}U_A^\dagger,\  \forall x,y \ \text{ and } \
\ \tilde{F}_{xy}=U_BF_{xy}U_B^\dagger,\ \forall x,y.
\ee
It is easy to verify  that   the matrices   $\{\tilde{E}_{xy}\}_y$  and $\{\tilde{F}_{xy}\}_y$ form valid quantum measurements and 
$$\tilde{\psi}=\sum_{i=1}^{d'}\lambda_i\ e_i\otimes e_i\in \C^{d'}\otimes \C^{d'},$$
is a valid quantum state with full Schmidt rank. 
Furthermore, the quantum strategy corresponding to  $\tilde{\psi},$ $\{\tilde{E}_{xy}\}_y$  and $\{\tilde{F}_{xy}\}$ also generates the correlation $p(y,y'|x,x')$ given in \eqref{eq:qcorrelationform}.  Another useful consequence of this fact is  that the operator $\rho = \mat(\tilde{\psi})$ is a diagonal matrix with positive~entries. When considering quantum strategies for the isomorphism game, we will often say we are ``working in the Schmidt basis of the shared state $\psi$". By this we mean that we have implicitly performed the above transformation on the shared state and measurement operators of our strategy.

Lastly, we  introduce a useful mathematical tool. Let $\vect : \C^{d_1 \times d_2} \to~\C^{d_1} \otimes \C^{d_2}$  be the linear map that takes the matrix $uv^\dagger$ to $u \otimes \overline{v},$ where $\overline{v}$ denotes  the entrywise complex conjugate of $v$. In other words, the map $\vect$ creates a vector from a matrix by stacking (the transpose of) its rows on top of each other. Also, let $\mat : \C^{d_1} \otimes \C^{d_2} \to \C^{d_1 \times d_2}$ be the inverse of the vectorization map. It is not hard to see that the $\vect$ map is an isometry, {\em i.e.,}
\be\label{eq:isometry}
\vect(A)^\dagger \vect(B) = \tr(A^\dagger B), \text{ for all } A,B.
\ee
Setting $\rho = \mat(\psi)$ we have that 
\begin{equation}\label{eqn:probs}
\psi^\dagger \left(E \otimes F\right) \psi = \vect(\rho)^\dagger \left(E \otimes F\right) \vect(\rho) = \vect(\rho)^\dagger \vect(E\rho F^\trans) = \tr(\rho^\dagger E \rho F^\trans),
\end{equation}
where we used \eqref{eq:isometry} and  the identity $\vect(AXB^\trans) = (A \otimes B) \vect(X),$
for Hermitian operators of appropriate size. This identity is crucial for our results in the next section.

\subsection{Characterizing perfect quantum strategies}\label{subsec:charqstrats}
In  Section~\ref{subsec:qstrats} we described the general form of a quantum strategy for the $(G,H)$-isomorphism game. In this section we  investigate quantum isomorphisms in more detail, and show that perfect  quantum strategies can always be chosen to take a specific form.

\begin{definition}
A nonlocal game is called \emph{synchronous} if the players share the same question set $X$,  the same answer set $Y$, and furthermore, 
\be\label{eq:synchronous}
V(y,y' | x, x) = 0, \text{ for all  } x\in X \text{ and } y\ne y'\in Y.
\ee 
 Analogously, a correlation is called  {\em synchronous} if  $p(y,y' | x, x) = 0,$ for all $ x\in X$ and $ y\ne y'\in Y$.
\end{definition}

For example, note that the $(G,H)$-isomorphism game is synchronous. Indeed, in this game  the  question and answer sets are   both equal to $V(G)\cup V(H)$. Furthermore, if the players are given the same vertex and they  respond with different vertices they lose.  This shows that  \eqref{eq:synchronous} is satisfied. 

Synchronous games have recently received  significant  attention in the literature due to the fact that their perfect  quantum strategies always have a special form. Specifically, the following result or a similar version  has appeared in various places~\cite{qhomos,RobersonThesis,MRV,Cameron07}.

\begin{lemma}\label{lem:synchgames}
Let  $\psi \in \C^d \otimes \C^d$, $\mathcal{E}_x = \{E_{xy} : y \in Y\},$ and $\mathcal{F}_x = \{F_{xy} : y \in Y\}$ for all $x \in X$ be a perfect quantum strategy for a synchronous game. If $\psi$ and the operators $E_{xy}$ and $F_{xy}$ are expressed in the Schmidt basis of $\psi$, and $\rho = \mat(\psi)$, then
\begin{enumerate}[(i)]
\item $E_{xy} = F_{xy}^\trans$ for all $x \in X$, $y \in Y$;
\item $E_{xy}$ and $F_{xy}$ are projectors for all $x \in X$, $y \in Y$;
\item $E_{xy}\rho = \rho E_{xy}$ and $F_{xy}\rho = \rho F_{xy}$ for all $x \in X$, $y \in Y$;
\item $p(y, y' | x, x')= \psi^\dagger\left(E_{xy} \otimes F_{x'y'}\right)\psi = 0\ $ if and only if $\  E_{xy}E_{x'y'} = 0$.
\end{enumerate}
\end{lemma}

Since the graph isomorphism game is synchronous, Lemma~\ref{lem:synchgames} applies to it. However, in our next result we 
show that even more conditions are met by perfect  quantum strategies for the isomorphism game.

\begin{theorem}\label{thm:stratform}Consider two graphs $G$ and $H$  and set $V = V(G) \cup V(H)$. Let  $\psi \in \C^d \otimes \C^d$, $\mathcal{E}_x = \{E_{xy} : y \in V\}, $ and  $\mathcal{F}_x = \{F_{xy} : y \in V\}$ for all  $x \in V$ be  a perfect  quantum strategy for the $(G,H)$-isomorphism game. If $\psi$ and the operators $E_{xy}$ and $F_{xy}$ are expressed in the Schmidt basis of $\psi$, and $\rho = \mat(\psi)$, then
\begin{enumerate}[(i)]
\item $E_{xy} = F_{xy}^\trans$ for all $x, y \in V$;
\item $E_{xy}$ and $F_{xy}$ are projectors for all $x, y \in V$;
\item $E_{xy}\rho = \rho E_{xy}$ and $F_{xy}\rho = \rho F_{xy}$ for all $x, y \in V$;
\item $p(y, y' | x, x') := \psi^\dagger\left(E_{xy} \otimes F_{x'y'}\right)\psi = 0$ if and only if $E_{xy}E_{x'y'} = 0$;
\item $E_{xy} = 0$ if $x,y \in V(G)$ or $x,y \in V(H)$;
\item $E_{xy} = E_{yx}$ for all $x, y \in V$.
\end{enumerate}
\end{theorem}
\proof
The first four conditions follow immediately from Lemma~\ref{lem:synchgames}. For $(v)$, consider  $g, g' \in V(G)$ and note that
\be\label{eq:csdfergrt}
0 = \sum_{y \in V} p(g', y| g, x) = \sum_{y \in V} \psi^\dagger\left(E_{gg'} \otimes F_{xy}\right)\psi = \psi^\dagger\left(E_{gg'} \otimes I\right)\psi = \tr(\rho \rho^\dagger E_{gg'}),
\ee
where the last equality follows from Equation~(\ref{eqn:probs}) and the cyclicity of the trace. Since we are working in the Schmidt basis of $\psi$, the matrix $\rho$ is diagonal with strictly positive diagonal entries. Therefore $\rho\rho^\dagger$ has full rank and thus it follows by \eqref{eq:csdfergrt} that $E_{gg'} = 0$. Similarly,  we have that $E_{hh'} = 0$ for all $h,h' \in V(H)$.

Lastly, we show $(vi)$. By the rules of the $(G,H)$-isomorphism game, we must have that $p(y, x'|x, y) = 0$ whenever  $x' \ne x$. Thus by~$(iv)$ we have that $E_{xy} E_{yx'} = 0$ for all $x' \ne x$. Therefore,
\[E_{xy} = E_{xy} \sum_{x'} E_{yx'} = E_{xy}E_{yx}.\]
Also, $E_{xy'}E_{yx} = 0$ for $y' \ne y$, and thus
\[E_{yx} = \sum_{y'} E_{xy'} E_{yx} = E_{xy}E_{yx}.\]
Combining the  two equations above  we get  that $E_{xy} = E_{yx}$.\qeds

\subsection{Two algebraic reformulations}
In  this section we use the structural properties of perfect quantum strategies to the $(G,H)$-isomorphism game we identified in Section~\ref{subsec:charqstrats}   to prove Result~\ref{thm:summary}.

First, we show the equivalence $(i)\Longleftrightarrow (ii)$ from Result~\ref{thm:summary}.

\begin{theorem}\label{thm:qreform}
Let $G$ and $H$ be graphs. Then $G \cong_q H$ if and only if there exist projectors $E_{gh}$ for $g \in V(G)$ and $h \in V(H)$ such that
\begin{enumerate}[(i)]
\item $\sum_{h \in V(H)} E_{gh} = I,\ $ for all $g \in V(G)$;
\item $\sum_{g \in V(G)} E_{gh} = I,\ $ for all $h \in V(H)$;
\item $E_{gh} E_{g'h'} = 0,\ $ if $\rel(g,g') \ne \rel(h,h')$.
\end{enumerate}
\end{theorem}
\proof
Using Theorem~\ref{thm:stratform}, it is relatively easy to see that Alice's operators $E_{gh}$ for $g \in V(G)$, $h \in V(H)$ from a perfect quantum strategy satisfy  Conditions $(i)$ $(ii)$ and $(iii)$.

Conversely, suppose that $E_{gh}$ for $g \in V(G)$ and $h \in V(H)$ satisfy the hypotheses of the theorem. Define $E_{hg} = E_{gh}$, and $E_{gg'} = E_{hh'} = 0$ for all for $h, h' \in V(H)$, $g, g' \in V(G)$. Furthermore, let $F_{xy} = E_{xy}^\trans$ for all $x, y \in V(G) \cup V(H)$. It is easy to see that $\mathcal{E}_x = \{E_{xy} : y \in V(G) \cup V(H)\}$ is a valid measurement for all $x \in V(G) \cup V(H)$, and similarly for $\mathcal{F}_x = \{F_{xy} : y \in V(G) \cup V(H)\}$.  Consider the quantum strategy where Alice and Bob respectively use the measurements  $\mathcal{E}_x$ and $\mathcal{F}_x$  on a shared  maximally entangled state $\psi_d= \frac{1}{\sqrt{d}} \sum_{i=1}^d e_i \otimes e_i$. By \eqref{eq:maxent} we  have that
\[p(y, y' | x, x') = \psi_d^\dagger \left(E_{xy} \otimes F_{x'y'} \right) \psi_d = \frac{1}{d}\tr(E_{xy}E_{x'y'}),\]
for all $x, x', y, y' \in V(G) \cup V(H)$. This fact combined with Condition $(iii)$ shows   that this is a perfect  strategy for the $(G,H)$-isomorphism game.\qeds

Recall that an alternative characterization of graph isomorphism is given by the equation $A_GP = PA_H$ for some permutation matrix $P$, where $A_G$ and $A_H$ are the adjacency matrices of two graphs. It turns out that one can use Theorem~\ref{thm:qreform} to obtain an analogous formulation of quantum graph isomorphism. 
First we will need the following definition:

\begin{definition}  
A  matrix $\P\in\bmat$ is called a \emph{projective permutation matrix of block size $d$} if it is unitary and all of its blocks are projectors.
\end{definition}

Note that a projective permutation matrix of block size one is a unitary matrix whose entries square to themselves, \emph{i.e.}, a permutation matrix. The following lemma shows that projective permutation matrices can be built out of projectors satisfying the first two conditions of Theorem~\ref{thm:qreform}. 

\begin{lemma}\label{lem:projperm2}
 A  matrix $\P=[[E_{ij}]]\in \bmat$ is a projective permutation matrix if and only if the matrix $E_{ij}$ is a projector for all $i,j \in [n]$ and 
\begin{enumerate}[(i)]
\item $\sum_{j=1}^n E_{ij} = I,$ for all $i \in [n]$;
\item $\sum_{i=1}^n E_{ij} = I,$ for all $j \in [n]$.
\end{enumerate}
\end{lemma}
\proof
It suffices that show that, assuming the matrices $\{E_{ij}\}_{ij}$ are projectors, Conditions $(i)$ and~$(ii)$  in the statement of the lemma  are equivalent to $\mathcal{P}$ being unitary.

First, assume that  Conditions $(i)$ and $(ii)$  hold. Since the matrices $\{E_{ij}\}_{ij}$ are projectors it follows that $E_{ik}E_{jk} = 0,$ for all $i \ne j\in [n]$.  This implies
\[\left(\mathcal{P}\mathcal{P}^\dagger\right)_{i,j} = \sum_{k} E_{ik}E_{jk} = \begin{cases}
\sum_k E_{ik}^2 = I, & \text{if } i = j \\
0, & \text{if } i \ne j.
\end{cases}
\]
Therefore $\mathcal{P}\mathcal{P}^\dagger = I$ and similarly we have that  $\mathcal{P}^\dagger\mathcal{P} = I$, \emph{i.e.}, $\mathcal{P}$ is unitary.

Conversely, suppose that $\mathcal{P}$ is unitary. Since  $\mathcal{P}\mathcal{P}^\dagger = I$ we have that   
\[I = \left(\mathcal{P}\mathcal{P}^\dagger\right)_{i,i} = \sum_{j} E^2_{ij} = \sum_{j} E_{ij}.\]
Analogously, using that $\mathcal{P}^\dagger \P=I,$ we get that $\sum_{i} E_{ij} = I$, and thus Conditions $(i)$ and $(ii)$ hold.\qeds

\begin{remark} 
In~\cite{qlatin}, Musto and Vicary introduced \emph{quantum Latin squares}. This is an $n \times n$ array of unit vectors in which each row and column forms an orthonormal basis of $\mathbb{C}^n$. They use quantum Latin squares  to construct  unitary error bases which are related to teleportation, dense coding, and quantum error correction. If $\mathcal{P}$ is a projective permutation matrix in which each projector $E_{ij}$ has rank one, then there exist unit vectors $\psi_{ij}$ such that $E_{ij} = \psi_{ij}\psi_{ij}^\dagger$. By Lemma~\ref{lem:projperm2} we have that $\psi_{ij}^\dagger\psi_{ij'} = 0$ when $j \ne j'$ and $\psi_{ij}^\dagger\psi_{i'j} = 0$ when $i \ne i'$. In other words, the vectors $\psi_{ij}$ form a quantum Latin square. Thus projective permutation matrices also generalize quantum Latin~squares.
\end{remark}

We are now ready to prove the equivalence $(i)\Longleftrightarrow (iii)$ from Result~\ref{thm:summary}.

\begin{theorem}\label{lem:projperm}
For any two graphs $G$ and $H$ we have that $G \cong_q H$ if and only if there exists a projective permutation matrix $\P\in \bmat$ (for some $d \in \mathbb{N}$) such that
\be\label{eq:generalizediso}
(A_G \otimes I_d)\mathcal{P} = \mathcal{P}(A_H \otimes I_d).
\ee
\end{theorem}
\proof
Let $\P=[[E_{gh}]]$ for $g \in V(G)$ and $h \in V(H)$. By Lemma~\ref{lem:projperm2}, the blocks  $\{E_{gh}\}$ are projectors and satisfy Conditions $(i)$ and $(ii)$ of Theorem~\ref{thm:qreform}. So we only need to show that, assuming these properties, the equation $(A_G \otimes I_d)\mathcal{P} = \mathcal{P}(A_H \otimes I_d)$ 
is equivalent to Condition~$(iii)$ of Theorem~\ref{thm:qreform}.
Note that $E_{gh}E_{g'h'} = 0$ whenever ($g = g'$ and $h \ne h'$) or ($h = h'$ and $g \ne g'$) is already guaranteed by Conditions $(i)$ and $(ii)$ of Lemma~\ref{lem:projperm2}. Thus, we only need to prove the remaining orthogonalities of Theorem~\ref{thm:qreform} $(iii)$.

The $(g,g')$-block of $A_G \otimes I_d$ is equal to  $I_d$ if $g$ and $g'$ are adjacent, and is $0$ otherwise. Similarly for the $(h,h')$-block of $A_H\otimes I_d$. Moreover, note  that for $g \in V(G)$ and $h \in V(H)$ we have 
\be\label{eq:expression}
\left((A_G \otimes I_d)\mathcal{P}\right)_{g,h} = \sum_{g': g' \sim g} E_{g'h},\  \text{ and }\  \left(\mathcal{P}(A_H\otimes I_d)\right)_{g,h} = \sum_{h': h' \sim h} E_{gh'}.
\ee
If Theorem~\ref{thm:qreform} $(iii)$ holds, then for all $g \in V(G)$ and $h \in V(H)$ we have  
\[\sum_{g' \sim g} E_{g'h} = \sum_{g' \sim g} E_{g'h}\sum_{h'} E_{gh'} = \sum_{g' \sim g} E_{g'h}\sum_{h' \sim h} E_{gh'} = \sum_{g'} E_{g'h} \sum_{h' \sim h} E_{gh'} = \sum_{h' \sim h} E_{gh'},\]
and therefore it follows by \eqref{eq:expression} that  $(A_G \otimes I_d)\mathcal{P} = \mathcal{P}(A_H \otimes I_d)$.

Conversely, if $(A_G\otimes I_d)\mathcal{P} = \mathcal{P}(A_H\otimes I_d)$, it follows by \eqref{eq:expression} that 
 \be\label{eq:sfefererge}
 \sum_{g': g' \sim g} E_{g'h} = \sum_{h': h' \sim h} E_{gh'}, \ \text{ for all } g \in V(G), h \in V(H).
 \ee 
 Furthermore, since the projectors $E_{gh}$  are mutually orthogonal for any fixed $h\in V(H)$ we have
 \be\label{eq:sfegrrtghrth}
 \left(\sum_{g': g' \sim g} E_{g'h}\right)^2 
= \sum_{g': g' \sim g} E_{g'h}, 
 \ee
and therefore, combining \eqref{eq:sfefererge} with \eqref{eq:sfegrrtghrth} it follows that 
\be\label{eq:sdvfergrtg}
\sum_{g' \sim g} E_{g'h} \sum_{h' \sim h} E_{gh'} 
= \sum_{g' \sim g} E_{g'h} 
= \sum_{g' \sim g} E_{g'h} \sum_{h'} E_{gh'}. 
\ee
As a consequence of \eqref{eq:sdvfergrtg} we get that 
\be\label{sdfdegrgtr}
\sum_{g' \sim g} E_{g'h} \sum_{h' \not{\sim} h} E_{gh'} = 0.
\ee
Taking traces in \eqref{sdfdegrgtr} we have
\be\label{eq:sdfsdegrt}
\sum_{g': g' \sim g, \ h': h' \not{\sim} h} \tr(E_{g'h}E_{gh'}) = 0.
\ee
Since  the matrices  $E_{gh}$ are positive semedefinite (as they are projectors), all the terms in \eqref{eq:sdfsdegrt}  must be nonnegative.  Therefore, we have  that $\tr(E_{g'h}E_{gh'}) = 0$ for all $g' \sim g$ and $h' \not\sim h$, which implies that $E_{g'h}E_{gh'} = 0$ for all $g' \sim g$ and $h' \not\sim h$. One can similarly show that $E_{g'h}E_{gh'} = 0$ if $h \sim h'$ and $g \not\sim g'$. So if one of $\rel(g,g')$ and $\rel(h,h')$ is ``adjacency" and the other is not, we have the desired orthogonalities. We also already noted at the beginning of the proof that when one of $\rel(g,g')$ and $\rel(h,h')$ is ``equality" and the other is not, we have the required  orthogonality. The only thing remaining is when one of $\rel(g,g')$ and $\rel(h,h')$ is ``distinct non-adjacency" and the other is not. However this is implied by what we already have, since the relationship which is not ``distinct non-adjacency" will be one of ``equality" or ``adjacency".\qeds

\begin{remark}
The above lemma shows that projective permutation matrices play the role of permutation matrices for quantum isomorphisms. In fact, just as any permutation matrix corresponds to an isomorphism from a complete (or empty) graph to itself, any projective permutation matrix corresponds to a quantum isomorphism from a complete (or empty) graph to itself.
\end{remark}

Since a projective permutation matrix is unitary, the equation $(A_G \otimes I_d)\mathcal{P} = \mathcal{P}(A_H \otimes I_d)$ is equivalent to $\mathcal{P}^\dagger(A_G \otimes I_d)\mathcal{P} = (A_H \otimes I_d)$, which implies that $A_G \otimes I_d$ and $A_H \otimes I_d$ have the same multiset of eigenvalues. Of course this means that $A_G$ and $A_H$ have the same multiset of eigenvalues, and thus quantum isomorphic graphs are cospectral with respect to their adjacency matrices. Since two graphs are quantum isomorphic if and only if their complements are, we have the following corollary:

\begin{corollary}
If $G\cong_q H$ then $G$ and $H$ are cospectral with cospectral complements.
\end{corollary}

Note that this is not the case for non-signalling/fractional isomorphism. Indeed, any two $n$-vertex, $k$-regular graphs are fractionally isomorphic but there are many such pairs that are not cospectral. From this it follows that quantum and non-signalling isomorphism are different:

\begin{corollary}
There exist graphs that are non-signalling isomorphic but not quantum isomorphic.
\end{corollary}

In an upcoming work~\cite{qiso2} we show that cospectrality is a consequence of a semidefinite relaxation of quantum isomorphism that we call $\psd$-isomorphism. This relation is strictly weaker than quantum isomorphism, but still stronger than non-signalling isomorphism.

\subsection{Quantum commuting isomorphisms}\label{subsec:qcisos}

We note here that there are other mathematical models for performing joint quantum measurements, and thus for playing nonlocal games, which are slightly different than the finite dimensional tensor product framework we have discussed so far. Firstly, one can consider allowing infinite dimensional Hilbert spaces to model the quantum systems of the players. In this case, the strategies are the same and the probabilities are given by the same expression, but the shared state $\psi$ and the operators $E_{xy}$ and $F_{xy}$ for Alice and Bob are allowed to be infinite dimensional. In general, it is not known whether allowing infinite dimensional spaces can allow one to win a nonlocal game perfectly when one cannot using finite dimensional strategies. However, though it is not obvious, it follows from results in~\cite{clevemittal} that these two models for quantum strategies are equivalent for the isomorphism game, as well as all other games we consider in this work. But there is yet another model for joint quantum measurements that we will see is different from the finite dimensional tensor product framework which we have focused on so far. We explain this model below.

In the tensor product framework, each party has their own (finite dimensional) Hilbert space that they act on with positive operators. In the \emph{quantum commuting framework}, both players share some, potentially infinite dimensional, Hilbert space $H$ on which they both act with positive elements of the space of bounded linear operators on $H$, denoted $\mathcal{B}(H)$. However, it is required that all of Alice's measurement operators commute with all of Bob's measurement operators. Thus if Alice performs the measurement $\{E_i\in \mathcal{B}(H)_+ : i\in [m_1]\}$ and Bob performs the measurement $\{F_j\in \mathcal{B}(H)_+ : j\in [m_2]\}$ on their shared state $\psi \in H$, then it is required that $E_iF_j = F_jE_i$ for all $i \in [m_1]$, $j \in [m_2]$, and the probability that they obtain outcome $(i,j)$ is given by $\langle E_iF_j \psi, \psi \rangle$.

The quantum commuting framework is more general than the tensor product framework given above. To see this note that if $\{E_i\in \mch_+^{d_1} : i\in [m_1]\}$ and $\{F_j\in \mch_+^{d_2} : j\in [m_2]\}$ are measurements used by Alice and Bob in the tensor product framework, then the measurements $\{E_i \otimes I_{d_2} \in \mch_+^{d_1} \otimes \mch_+^{d_2} : i\in [m_1]\}$ and $\{I_{d_1} \otimes F_j\in \mch_+^{d_1} \otimes \mch_+^{d_2} : j\in [m_2]\}$ are valid joint measurements in the quantum commuting framework that result in the same outcome probabilities. It is also known, though it is highly nontrivial, that when restricted to finite dimensional Hilbert spaces, the two frameworks are equivalent~\cite{tsirelson06}. Thus we always allow for infinite dimensional Hilbert spaces when considering the quantum commuting framework. It was only recently shown by Slofstra~\cite{slofstra16} that the quantum commuting framework can allow one to win some nonlocal games that cannot be perfectly won using the tensor product framework. In Section~\ref{sec:separation} we will use his result to show that this also holds in the specific case of isomorphism games.

In light of the above, we can define two graphs $G$ and $H$ to be \emph{quantum computing isomorphic}, denoted $G \cong_{qc} H$, whenever there exists a perfect quantum commuting strategy for the $(G,H)$-isomorphism game. Such strategies for the graph coloring game were investigated in~\cite{paulsen13} and~\cite{paulsen14}. The analysis in the latter applies to any synchronous game, and thus from their results we can obtain the following:

\begin{lemma}\label{lem:synchstrats}
Consider a synchronous game with input sets $X$, output sets $Y$, and verifcation function $V$. There exists a perfect quantum commuting strategy for this game if and only if there exists a unital $C^*$-algebra $\mathcal{A}$, a faithful tracial state $s: \mathcal{A} \to \mathbb{C}$, and projections $E_{xy} \in \mathcal{A}$ for $(x,y) \in X \times Y$ such that
\begin{enumerate}
\item $\sum_{y \in Y} E_{xy} = I;$
\item $s(E_{xy}E_{x'y'}) = 0$ if $V(y,y'|x,x') = 0$.
\end{enumerate}
\end{lemma}
Note that \emph{tracial state} on a unital $C^*$-algebra $\mathcal{A}$ is a linear functional $s: \mathcal{A} \to \mathbb{C}$ such that $s(I) = 1$, $s(A^*A) \ge 0$ for all $A \in \mathcal{A}$, and $s(AB) = s(BA)$ for all $A,B \in \mathcal{A}$. The tracial state $s$ is \emph{faithful} if $s(A^*A) = 0$ if and only if $A = 0$. Note that if $A$ and $B$ are projections, then $s(AB) = 0$ implies that $AB = 0$ just like in the finite dimensional case (as long as $s$ is faithful). For the finite dimensional matrix algebra $\mathbb{C}^{d \times d}$, there is a unique tracial state given by $s(M) = \tr(M)/d$.

Using the above, we can prove the following analog of Theorem~\ref{thm:qreform}. We omit the proof since it is similar to that of Theorem~\ref{thm:qreform}.
\begin{theorem}\label{thm:qcreform}
Let $G$ and $H$ be graphs. Then $G \cong_{qc} H$ if and only if there exists a $C^*$-algebra $\mathcal{A}$ which admits a faithful tracial state, and projections $E_{gh} \in \mathcal{A}$ for $g \in V(G)$ and $h \in V(H)$ such that
\begin{enumerate}[(i)]
\item $\sum_{h \in V(H)} E_{gh} = I,\ $ for all $g \in V(G)$;
\item $\sum_{g \in V(G)} E_{gh} = I,\ $ for all $h \in V(H)$;
\item $E_{gh} E_{g'h'} = 0,\ $ if $\rel(g,g') \ne \rel(h,h')$.
\end{enumerate}
\end{theorem}

Note that since quantum commuting strategies for nonlocal games are more general than quantum tensor product strategies, we have that two graphs being quantum isomorphic implies that they are also quantum commuting isomorphic. Similarly, since quantum commuting strategies are also non-signalling, we have that any two quantum commuting isomorphic graphs are non-signalling isomorphic. In summary, we have that for any two graphs $G$ and $H$
\begin{equation}\label{eqn:imps}
G \cong H \ \Rightarrow \ G \cong_q H \ \Rightarrow \ G \cong_{qc} H \ \Rightarrow \ G \cong_{ns} H.
\end{equation}
By the end of this work we will see that all of these implications are strict, i.e., none of the four relations above are equivalent.

\subsection{Necessary conditions from quantum homomorphisms}

A \emph{homomorphism} from $G$ to $H$ is an adjacency preserving function $\varphi: V(G) \to V(H)$, \emph{i.e.}, if $g \sim g'$ then $\varphi(g) \sim \varphi(g')$. When such a function exists, we write $G \to H$. In~\cite{qhomos}, the homomorphism game was introduced and with it the notion of quantum homomorphisms. This was in fact part of the initial inspiration for the isomorphism game and quantum isomorphisms. In this section we will see that, as in the classical case, any pair of quantum isomorphic graphs must admit quantum homomorphisms between each other in both directions. This will imply that quantum isomorphic graphs must have equal values for several \emph{quantum parameters} such as the quantum chromatic number, thus providing us with many necessary conditions for a pair of graphs to be quantum~isomorphic.

Given graphs $G$ and $H$, the $(G,H)$-homomorphism game is played as follows: Alice and Bob are given vertices $\ga, \gb \in V(G)$ and must respond with vertices $\ha, \hb \in V(H)$ respectively. If $\ga = \gb$, then to win they must satisfy $\ha = \hb$, and if $\ga \sim \gb$ then they must satisfy $\ha \sim \hb$. Similarly to the isomorphism game, it is not difficult to show that classical players can win the $(G,H)$-homomorphism game perfectly  if and only if there exists a homomorphism from $G$ to $H$. Motivated by this, in~\cite{qhomos} they say that there is a quantum homomorphism from $G$ to $H$, and write $G \qarrow H$ if there exists a perfect quantum strategy for the $(G,H)$-homomorphism~game.

Suppose that Alice and Bob have a perfect strategy for the $(G,H)$-isomorphism game. If we restrict their possible inputs to only the vertices from $G$, then it is easy to see that they will always satisfy the winning conditions of the $(G,H)$-homomorphism game: if they are give the same vertices from $G$ they will respond with the same vertices from $H$ and if they are given adjacent vertices of $G$ they will respond with adjacent vertices of $H$. Therefore, if two players have a perfect (classical or quantum) strategy for the $(G,H)$-isomorphism game, then they can use the same strategy, but restricted to inputs from $V(G)$, to perfectly win the $(G,H)$-homomorphism game. Thus we have the following:

\begin{lemma}\label{lem:qiso2qhomo}
If $G \cong_q H$, then $G \qarrow H$, $H \qarrow G$, $\overline{G} \qarrow \overline{H}$, and $\overline{H} \qarrow \overline{G}$.
\end{lemma}

The usefulness of the above is that we can combine it with known results relating quantum homomorphisms and certain quantum analogs of classical graph parameters. For instance, the \emph{quantum chromatic number} of $G$, denoted $\chi_q(G)$, is defined as the minimum $c \in \mathbb{N}$ such that $G \qarrow K_c$, where $K_c$ is the complete graph on $c$ vertices. It follows essentially from the definition (and the fact that quantum homomorphisms can be composed~\cite{qhomos}), that if $G \qarrow H$ then $\chi_q(G) \le \chi_q(H)$. This property of $\chi_q$ is known as being \emph{quantum homomorphism monotone}, and it is analogous to a similar statement for chromatic number and classical homomorphisms. By the above lemma, this shows that if $G \cong_q H$, then $\chi_q(G) = \chi_q(H)$, and similarly for the complements. There are other quantum parameters defined similarly to $\chi_q$, such as the quantum clique number $\omega_q(G) := \min\{c : K_c \qarrow G\}$, or the quantum independence number $\alpha_q(G) := \omega_q(\overline{G})$. Similarly to $\chi_q$, these parameters are  equal for quantum isomorphic graphs.

For the above examples of quantum graph parameters, proving quantum homomorphism monotonicity is straightforward, since the parameters themselves are defined in terms of quantum homomorphisms. However, there are some interesting examples of graph parameters that are not defined in this way, but still turn out to be quantum homomorphism monotone. For instance, the well known Lov\'{a}sz theta number (of the complement) was proven to be quantum homomorphism monotone in~\cite{qhomos}, as were two variants by Schrijver and Szegedy in~\cite{RobersonThesis}. For us, there are two other, lesser known, quantum homomorphism monotone parameters that will be important for this work. We introduce both below.

A \emph{$(d/r)$-projective representation} of a graph $G$ is an assignment of $d \times d$ projectors of rank $r$ to the vertices of $G$ such that projectors assigned to adjacent vertices are orthogonal. The \emph{projective rank} of a graph $G$, denoted $\xi_f(G)$, is the infimum of $\frac{d}{r}$ such that $G$ admits a $(d/r)$-projective representation.

A \emph{projective packing} of a graph $G$ is an assignment, $g \mapsto E_g \in \mathbb{C}^{d \times d}$, of $d \times d$ projectors for some $d \in \mathbb{N}$ such that adjacent vertices receive orthogonal projectors. Note that there is no uniformity condition on the rank of the projectors as there is for a projective representation. The {\em value} of a projective packing is equal to $\frac{1}{d}\sum_{g \in V(G)}\rk(E_g)$, and the {\em projective packing number} of $G$, denoted $\alpha_p(G)$, is the supremum of the values of all projective packings of $G$.

If a graph $G$ can be $c$-colored, then by replacing color $i$ with the projection onto the $i^\text{th}$ standard basis vector in $\mathbb{C}^c$ gives a $(c/1)$-projective representation of $G$, and thus $\xi_f(G) \le \chi(G)$. The inequality can be strict, and in fact the projective rank also lower bounds both the fractional and quantum chromatic numbers. In a sense, the projective rank can be thought of as a fractional quantum chromatic number. It was shown in~\cite{qhomos} that $\xi_f$ is quantum homomorphism monotone, and moreover that if $G$ has a quantum homomorphism to $H$ and the latter has a $(d/r)$-projective representation then $G$ has a $(d'/r')$-projective representation for some $d',r'$ such that $\frac{d'}{r'} = \frac{d}{r}$.

Similarly, if $S \subseteq V(G)$ is an independent set of vertices, then assigning the identity matrix to all vertices of $S$ and the zero matrix to all other vertices produces a projective packing of $G$ with value $|S|$. Therefore $\alpha(G) \le \alpha_p(G)$, and in fact $\alpha_q(G) \le \alpha_p(G)$ holds. It was shown in~\cite{RobersonThesis} that $G \qarrow H$ implies that $\alpha_p(\overline{G}) \le \alpha_p(\overline{H})$, and moreover if $\overline{G}$ has a projective packing of value $\gamma$ then $\overline{H}$ has a projective packing of value $\gamma$.

By the above discussion, we have the following:

\begin{lemma}\label{lem:projrank}
If $G\cong_qH$ then $\xi_f(G) = \xi_f(H)$ and $\xi_f(\overline{G}) = \xi_f(\overline{H})$. Moreover, if $G$ has a $(d/r)$-projective representation then $H$ has a $(d'/r')$-projective representation where $\frac{d'}{r'} = \frac{d}{r}$.
\end{lemma}
Similarly, we have:
\begin{lemma}\label{lem:projpack}
If $G\cong_qH$ then $\alpha_p(G) = \alpha_p(H)$ and $\alpha_p(\overline{G}) = \alpha_p(\overline{H})$. Moreover, if $G$ has a projective packing of value $\gamma$ then $H$ has a projective packing of value~$\gamma$.
\end{lemma}

In our upcoming work~\cite{qiso2}, we use Lemma~\ref{lem:projrank} above to show that quantum isomorphism and one of our semidefinite relaxations of quantum isomorphism are indeed different relations. In this work we will use Lemma~\ref{lem:projpack} for our reduction of linear binary constraint system games to isomorphism games in Section~\ref{subsec:reduction}. We note that the projective packing number is similar in many ways to the quantum independence number. In fact, there are no graphs $G$ for which it is known that $\alpha_q(G) \ne \lfloor \alpha_p(G) \rfloor$. Moreover, the following was shown in~\cite{MRV}:

\begin{lemma}\label{lem:projpackequal}
Let $G$ be a graph. Then $\alpha_p(G) \le \chi(\overline{G})$, and there exists a projective packing of $G$ of value $\chi(\overline{G})$ if and only if $\alpha_q(G) = \chi(\overline{G})$.
\end{lemma}

For our results on quantum commuting isomorphism, we will need an analog of projective packings that allows the projections we assign to our vertices to be more general objects. Such an analog of projective representations/projective rank was given in~\cite{paulsen14}, and we can adapt their approach here. Let $\mathcal{A}$ be a unital $C^*$-algebra that admits a faithful tracial state $s$. An assignment, $g \mapsto E_g \in \mathcal{A}$ , of projections from $\mathcal{A}$ to the vertices of a graph $G$ is a \emph{tracial packing} of $G$ if $E_gE_{g'} = 0$ whenever $g \sim g'$. The value of a such a tracial packing is $\sum_{g \in V(G)} s(E_g) = s(\sum_{g \in V(G)} E_g)$. The \emph{tracial packing number} of $G$, denoted $\alpha_{tr}(G)$, is the supremum of values of tracial packings of $G$.

One can also define quantum commuting homomorphisms and thus the quantum commuting independence number, denoted $\alpha_{qc}$ analogously to quantum homomorphisms and the quantum independence number above. Also, it is not difficult to adapt the proofs of Lemmas~\ref{lem:qiso2qhomo}, \ref{lem:projpack}, and~\ref{lem:projpackequal} to obtain analogs of these results for quantum commuting iso/homomorphisms and tracial packing number. We note that quantum commuting homomorphisms have been investigated in~\cite{paulsen14} and~\cite{ortiz}.

\section{Separating isomorphism and quantum isomorphism}\label{sec:separation}
In this section we prove Result~\ref{thm:qvsclassicalseparation}, \emph{i.e.}, we  construct pairs of graphs that are quantum isomorphic but not isomorphic. For this we  introduce a type of game investigated by Cleve and Mittal~\cite{clevemittal} known as \emph{binary constraint system} (BCS) games. We will show that, in the linear case, one can reduce the existence of a perfect classical (quantum) strategy for a BCS game to the existence of a perfect classical (quantum) strategy to a corresponding isomorphism~game.

\subsection{Binary constraint systems games}\label{subsec:BCSgames}

A linear binary constraint system (BCS) $\mathcal{F}$ consists of a family of binary variables $x_1, \ldots, x_n$ and  {\em constraints} $C_1, \ldots, C_m$, where each $C_\ell$ is a linear equation over $\F_2$ in some subset of the variables. Thus $C_\ell$ takes the form $\sum_{x_i \in S_\ell} x_i = b_\ell$ for some $S_\ell \subseteq \{x_1, \ldots, x_n\}$ and $b_\ell \in \{ 0,1\}$. We say that a BCS is \emph{satisfiable} if there is an assignment of values from $\F_2$ to the variables $x_i$ such that every constraint $C_\ell$ is satisfied. Such an assignment is known as a \emph{satisfying assignment}.

An example of a linear BCS is the following:
\begin{align}\label{eqn:BCS}
x_1 + x_2 + x_3 &= 0 \quad &x_1 + x_4 + x_7 = 0 \nonumber\\
x_4 + x_5 + x_6 &= 0 \quad &x_2 + x_5 + x_8 = 0 \\
x_7 + x_8 + x_9 &= 0 \quad &x_3 + x_6 + x_9 = 1\nonumber
\end{align}
where addition is over $\F_2$.
Note that  the BCS given above is not satisfiable.  Indeed, every variable appears in exactly two constraints and thus summing up all equations modulo 2 we get $0=1$.

To any  linear BCS $\mathcal{F}$ we associate  the following nonlocal game, which we call the {\em BCS game}. In the BCS game, the verifier gives  Alice a constraint $C_\ell$ and Bob  a constraint $C_k$. In order to win, they must each respond with an assignment of values to the variables in their respective constraints such that those constraints are satisfied. Furthermore, for the variables in $S_\ell \cap S_k$, Alice and Bob must agree on their assignment. Note that if they are given the same constraint, these conditions imply that they must give the same~response.

As with the other nonlocal games we have considered in this work, it is not difficult to see that
 Alice and Bob can win the BCS game  classically with probability 1 if and only if the corresponding BCS is satisfiable. This motivates the following definition. 

\begin{definition}
A linear BCS is called \emph{quantum satisfiable} if there exists a perfect quantum strategy for the corresponding BCS game.
\end{definition}

We note that  Cleve and Mittal in~\cite{clevemittal} also define a nonlocal game corresponding to a linear BCS that admits a perfect classical strategy if and only if the underlying BCS is satisfiable.  However,   their construction is slightly different than the game we devise  here. Specifically, in their game, the verifier gives   Alice  a  constraint $C_\ell$  and to  Bob   a single variable $x_i\in C_\ell$.  Alice returns an assignment for the variables in her constraint and Bob an assignment for his variable.  The winning conditions are  that $(i)$ Alice's assignment must satisfy $C_\ell$ and $(ii)$  Bob's assignment on $x_i$ must  be consistent with Alice's assignment. Moreover, they define quantum satisfiability in terms of something they call a quantum satisfying assignment for the equations in the BCS. However, their main result is that a BCS game has a perfect quantum strategy if and only if the BCS has a quantum satisfying assignment, and from this it easily follows that a perfect quantum strategy exists for our version of the BCS game if and only if one exists for theirs. Moreover, this implies that our notion of quantum satisfiability is the same as theirs. In the case of quantum commuting strategies, the equivalence of the two versions of the BCS game follow from the results in~\cite{cleveslofstra}. The reason we define the game differently is simply because the reduction to quantum isomorphism is more natural for this version.

There are many classes of linear BCS's that are quantum satisfiable but not satisfiable. Indeed, the example given above in~(\ref{eqn:BCS}) corresponds to the Mermin-Peres magic square game~\cite{mermin} which has a perfect quantum strategy. One can also use a result of Arkhipov~\cite{arkhipov} to construct a linear BCS that is quantum satisfiable but not satisfiable from any non-planar graph.

\subsection{The Reduction}\label{subsec:reduction}

In this section we   prove that (quantum) satisfiability of a linear BCS can be reduced to (quantum) graph isomorphism. As a first step we introduce the graphs  we use in the reduction.

To any linear BCS $\FF$ with $m$ constraints we associate  the  graph $\GF$ which is defined as  follows: For each constraint $C_\ell$, and each assignment $f: S_\ell \to \F_2$ that \emph{satisfies} $C_\ell$ we include a  vertex~$(\ell, f)$. Furthermore, we add an  edge between two vertices $(\ell, f)$ and $(k, f')$ if they are \emph{inconsistent}, \emph{i.e.}, if there exists $x_i \in S_\ell \cap S_k$ such that $f(x_i) \ne f'(x_i)$. We remark that this construction is related to the FGLSS reduction from~\cite{FGLSS}, which is well known in approximability literature.

Note that all  vertices of $\GF$  corresponding to a fixed constraint are pairwise adjacent. Thus,  for any linear BCS $\FF$, any independent set  in $\GF$ contains at most one vertex corresponding to each constraint. Therefore $\alpha(\GF) \le m$ for any linear BCS $\FF$ with $m$ constraints.

Given any linear BCS $\FF$, we define the {\em homogenization} of $\FF$, denoted by $\FF_0$, to be the linear BCS obtained from $\FF$ by changing the righthand sides of all of the constraints to 0. Note that the homogenization a linear BCS always has a solution, namely the all-zero assignment. Also note that $\GF$ and $\GFO$ have the same number of vertices.
 
We now show that   $\GFO$ always contains an independent set of size $m$ (and thus $\alpha(\GFO)=m$).   For each constraint $C_\ell$, let $f^0_\ell$ denote the zero assignment to the variables in $S_\ell$. Note that $(\ell, f^0_\ell)$ is a vertex of $\GFO$. Moreover, $(\ell,f^0_\ell)$ and $(k,f^0_k)$ are not adjacent in $\GFO$ since $f^0_\ell$ and $f^0_k$ are just restrictions of the same function (the zero assignment to all variables) and thus they necessarily agree on the intersection of their domains. Thus the vertices $\{(\ell, f^0_\ell) : \ell \in [m]\}$ form an independent set in $\GFO$ of size $m$. Therefore, for any linear BCS $\FF$ with $m$ constraints we have
 \be\label{eq:alphagfo} 
 \alpha(\GFO) = m.
\ee

We are now ready to  prove that satisfiability of a linear BCS $\FF$  can be reduced to deciding whether 
 $\GF$ and $\GFO$ are isomorphic.

\begin{theorem}\label{thm:BCS2iso}
Let $\FF$ be a linear BCS with $m$ constraints. Then the following are equivalent:
\begin{enumerate}[(i)]
\item $\FF$ is satisfiable;
\item The graphs $\GF$ and $\GFO$ are isomorphic;
\item $\alpha(\GF) = m$. 
\end{enumerate}
\end{theorem}
\proof
$(i)\Longrightarrow (ii).$ Suppose that $\FF$ is satisfiable and let  $F : \{x_1, \ldots, x_n\} \to \F_2$ be  a satisfying assignment. For each constraint $C_\ell$, let $F_\ell$ be the restriction of $F$ to the set $S_\ell$. Define a function $\varphi: V(\GF) \to V(\GFO)$ as follows. For each vertex $(\ell, f)$ of $\GF$, set  $\varphi(\ell,f) = (\ell, f \oplus F_\ell)$, where $f \oplus F_\ell$ is defined to be the function from $S_\ell$ to $\F_2$ given by $(f \oplus F_\ell)(x_i) = f(x_i) + F_\ell(x_i)$. 

We first show  that $\varphi$  is a function to the vertices of $\GFO$. For a vertex $(\ell, f) \in V(\GF)$, the constraint $C_\ell$ has the form $\sum_{x_i \in S_\ell} x_i = b_\ell$. By assumption, both $f$ and $F_\ell$ satisfy $C_\ell$, {\em i.e.,} we have that $\sum_{x_i \in S_\ell} f(x_i) = b_\ell$ and $\sum_{x_i \in S_\ell} F_\ell(x_i) = b_\ell$. Adding these up we get $\sum_{x_i \in S_\ell} (f \oplus F_\ell)(x_i) = b_\ell + b_\ell = 0$ and so $(\ell, f \oplus F_\ell)$ is indeed a vertex of $\GFO$. It is also easy to see that $\varphi$ is an injection and therefore also a bijection.

Next we  show that $\varphi$ preserves adjacency. Suppose that $(\ell, f)$ and $(k, f')$ are adjacent in~$\GF$. Then there exists $x_i \in S_\ell \cap S_k$ such that $f(x_i) \ne f'(x_i)$. It is easy to see that $f(x_i) + F_\ell(x_i) = f(x_i) + F(x_i) \ne f'(x_i) + F(x_i) = f'(x_i) + F_k(x_i)$ and thus $(\ell, f \oplus F_\ell)$ is adjacent to $(k, f' \oplus F_k)$ in $\GFO$. So $\varphi$ preserves adjacency and the proof that it preserves non-adjacency is~similar. This implies that $\varphi$ is an isomorphism and thus $\GF$ and $\GFO$ are isomorphic.

\medskip 

$(ii)\Longrightarrow (iii).$ We have already seen  in \eqref{eq:alphagfo} that for any linear BCS $\FF$ with $m$ constraints we have that  $\alpha(\GFO) = m$. By assumption  $\GF\cong \GFO$ and the claim follows.

\medskip 

$(iii)\Longrightarrow (i).$  Finally, suppose that $\alpha(\GF) = m$ and that $T$ is an independent set meeting this bound. As all  vertices of $\GF$  corresponding to a fixed constraint are pairwise adjacent, we must have that $T$ contains a unique vertex of the form $(\ell, f)$ for every $\ell \in [m]$. Therefore, we can define $f_\ell : S_\ell \to \F_2$ to be such that $(\ell, f_\ell) \in T$ for all $\ell \in [m]$. We will use these partial assignments to define a satisfying assignment $F$ for the BCS~$\FF$. Consider a variable $x_i$ and let $\ell\in  [m]$ such that $x_i \in S_\ell$. We define $F(x_i) = f_\ell(x_i)$. It remains  to show that $F$ is well-defined. Since $T$ is an independent set in $\GF$, if $k \ne \ell$ and $x_i \in S_k$, we must have that $f_\ell(x_i) = f_k(x_i)$. Therefore, the restriction of $F$ to $S_k$ is equal to $f_k$ for all $k \in [m]$. This implies that $F$ satisfies all of the constraints and is therefore a satisfying assignment.\qeds

Next we prove the quantum analog of Theorem~\ref{thm:BCS2iso}:

\begin{theorem}\label{thm:qBCS2qiso}
Let $\FF$ be a linear BCS with $m$ constraints. Then the following are equivalent:
\begin{enumerate}[(i)]
\item $\FF$ is quantum satisfiable;
\item The graphs $\GF$ and $\GFO$ are quantum isomorphic;
\item There exists a projective packing of $\GF$ of value $m$;
\item $\alpha_q(\GF) = m$.
\end{enumerate}
\end{theorem}
\proof

$(i)\Longrightarrow (ii).$ Suppose that $\FF$ is quantum satisfiable, \emph{i.e.}, there exists a perfect quantum strategy for the BCS game for $\FF$. We now describe a perfect strategy for the $(\GF, \GFO)$-isomorphism  game that uses  the perfect quantum strategy for the BCS game as a subroutine.

In the $(\GF, \GFO)$-isomorphism  game, Alice receives a vertex $(\ell_A,f_A) \in V(\GF) \cup V(\GFO)$. Upon receiving her question, she uses the perfect strategy for the BCS game to obtain an assignment $f'_A: S_{\ell_A} \to \F_2$ that satisfies $C_{\ell_A}$ in $\FF$, and responds with the vertex $(\ell, f_A \oplus f'_A)$. Similarly, Bob  receives a vertex $(k_B,f_B) \in V(\GF) \cup V(\GFO)$.  Using the perfect strategy for the BCS  game  he obtains an assignment $f'_B$  satisfying $C_{k_B}$ and he respond withs $(k_B,f_B \oplus f'_B)$. Note that $(f_A \oplus f'_A) \oplus f'_A = f_A$ and so without loss of generality we may assume that $(\ell_A, f_A), (k_B, f_B) \in V(\GF)$.

We now show that  the strategy for the $(\GF, \GFO)$-isomorphism game described above is perfect. For this, we need to show that  Conditions   \eqref{cond1}  and \eqref{cond2} are satisfied.   First, 
 note that since $f'_A$ satisfies constraint $C_{\ell_A}$ in $\FF$,  Alice's output $(\ell_A, f_A \oplus f'_A)$ is in $V(\GF)$ if  her input $(\ell_A, f_A)$ was in  $V(\GFO)$ and vice versa, thus Condition \eqref{cond1} of the isomorphism game is met.
Second, suppose that Alice and Bob's inputs were equal. Since they are using a perfect strategy for the BCS game,  the functions $f'_A$ and $f'_B$ are also the same and thus their outputs $(\ell_A, f_A \oplus f'_A)$ and $(k_B, f_B \oplus f'_B)$ are equal. Third, suppose that their inputs $(\ell_A, f_A)$ and $(k_B,f_B)$  were adjacent. By definition,   there exists $x_i \in S_\ell \cap S_k$ such that $f_A(x_i) \ne f_B(x_i)$. However, since they are using a perfect strategy for the BCS game, we have that $f'_A(x_i) = f'_B(x_i)$ and thus $(f_A \oplus f'_A)(x_i) \ne (f_B \oplus f'_B)(x_i)$. Therefore, their outputs $(\ell_A, f_A \oplus f'_A)$ and $(k_B, f_B \oplus f'_B)$ will be adjacent. Lastly, the proof that they output distinct non-adjacent vertices upon receiving distinct non-adjacent input vertices is similar. Therefore, Alice and Bob can win the $(\GF, \GFO)$-isomorphism game perfectly with this strategy. Since the strategy they used for the BCS game could be realized by quantum measurements on a shared entangled state, so can this one. Thus $\GF \cong_q \GFO$ and the proof is concluded.

\medskip 
 $(ii)\Longrightarrow (iii).$ Suppose that $\GF$ and $\GFO$ are quantum isomorphic. By \eqref{eq:alphagfo} we have that $\alpha(\GFO)=m$. This implies that $\GFO$ also has a projective packing of value $m$. Since $\GF \cong_q \GFO$, it follows  by Lemma~\ref{lem:projpack}  that $\GF$ must also have a projective packing of value $m$.

\medskip
$(iii)\Longrightarrow (i).$
Suppose that $\GF$ has a projective packing $(\ell, f) \mapsto E_{(\ell,f)} \in \mathbb{C}^{d \times d}$ of value $m$. Since the vertices corresponding to a single constraint form a clique, we have that the projectors assigned to those vertices are all mutually orthogonal. From this it follows that
\be\label{cevefv}
\sum_{f: (\ell, f) \in V(\GF)} \rk(E_{(\ell,f)}) \le d, \  \text{ for all } \ell \in [m].
\ee
Furthermore, we  have that
\be\label{eq:whatheverrrrr}
m = \frac{1}{d} \sum_{(\ell, f) \in V(\GF)} \rk(E_{(\ell,f)}) = \frac{1}{d} \sum_{\ell \in [m]} \sum_{f: (\ell, f) \in V(\GF)} \rk(E_{(\ell,f)}) \le \frac{1}{d} md = m,
\ee
where for the  inequality we used \eqref{cevefv}. Thus, equality holds throughout in \eqref{eq:whatheverrrrr}. 
This implies~that
$$\sum_{f: (\ell, f) \in V(\GF)} \rk(E_{(\ell,f)}) = d, \  \text{ for all } \ell \in [m],$$ 
which is possible  if and only if 
\be\label{eq:projmeasuremtn}
\sum_{f: (\ell, f) \in V(\GF)} E_{(\ell,f)} = I_d.
\ee
In view of \eqref{eq:projmeasuremtn}, the matrices  $\{E_{(\ell, f)} : f \text{ satisfies } C_\ell \}$ form   quantum measurement for each $\ell \in [m]$. 

To conclude the proof, we use these measurements to construct a perfect quantum strategy for the BCS game for~$\FF$. Specifically,  the players share the maximally entangled state $\psi_d  = \frac{1}{\sqrt{d}} \sum_{i=1}^d e_i \otimes e_i$. Upon  receiving constraint $C_\ell$, Alice   performs the measurement $\{E_{(\ell, f)} : f \text{ satisfies } C_\ell \}$ on her half of $\psi_d$ to obtain an assignment  $f: S_\ell \to \F_2$ that satisfies $C_\ell$. Upon receiving constraint $C_k$, Bob acts similarly, except that he performs the measurement $\{E^\trans_{(\ell, f)} : f \text{ satisfies } C_\ell \}$ to get an assignment  $f': S_k \to \F_2$ that satisfies $C_k$. The corresponding correlation is given by
\be\label{eq:correlationprojectors}
p(f,f'|C_\ell,C_k) = \psi_d^\dagger \left(E_{(\ell,f)} \otimes E^\trans_{(k,f')}\right)\psi_d=\frac{1}{d} \tr(E_{(\ell,f)}E_{(k,f')}),
\ee
where for the last equality we use \eqref{eq:maxent}.

It remains to check that this strategy wins the BCS game for $\FF$ perfectly. For this we need to show that the correlation defined in \eqref{eq:correlationprojectors} evaluates to zero  whenever the winning conditions of the BCS game are not fulfilled (recall Equation \eqref{eq:perfectsttategies}).
Now, by construction of the measurements, Alice and Bob  always output an assignment that satisfies their individual constraints. So the only thing to check is that the players  are consistent on any variables contained in both of their constraints. However, if there exists $x_i \in S_\ell \cap S_k$ such that $f(x_i) \ne f'(x_i)$, then the vertices $(\ell, f)$ and $(k,f')$ are adjacent in $\GF$. Therefore,  the projectors $E_{(\ell,f)}$ and $E_{(k,f')}$ are orthogonal since they originated  from a projective packing. As a consequence, it follows by \eqref{eq:correlationprojectors} that the probability of Alice and Bob responding with $f$ and $f'$ respectively upon being given constraints $C_\ell$ and $C_k$ is equal to zero.

\medskip 

$(iii) \Longrightarrow (iv)$. First, note that we can color the complement of $\GF$ with $m$ colors because the vertices corresponding to a fixed constraint of $\FF$ are an independent set in $\overline{\GF}$. Therefore, $\chi(\overline{\GF}) \le m$. Thus if there exists a projective packing of value $m$, by Lemma~\ref{lem:projpackequal} we have that $\chi(\overline{\GF}) = m$ and that $\alpha_q(\GF) = m$. 

\medskip 

$(iv) \Longrightarrow (iii)$.
Conversely, if $\alpha_q(\GF) = m$ then we must have that $\chi(\overline{\GF}) = m$. Then  Lemma~\ref{lem:projpackequal} implies that there exists a projective packing of value $m$.\qeds

As a corollary of the above two theorems, we have that isomorphism and quantum isomorphism are distinct relations on graphs:

\begin{theorem}\label{thm:result3}
There exists graphs that are quantum isomorphic but not isomorphic. In particular, if 
 $\FF$ is a linear BCS that is quantum satisfiable but not satisfiable, then the graphs $\GF$ and $\GFO$ are quantum isomorphic but not isomorphic.
\end{theorem}

The smallest example of a quantum satisfiable but not satisfiable linear BCS that we know of is the Mermin magic square BCS given in~(\ref{eqn:BCS}). The two corresponding graphs each have 24 vertices. Interestingly, both of the graphs have automorphism groups that act transitively on their vertices. In fact, both of the graphs are Cayley graphs. We present the two graphs in Figures~\ref{fig:qiso1} and~\ref{fig:qiso2} below.

\begin{figure}[h!]
\begin{center}
\includegraphics[scale=.52]{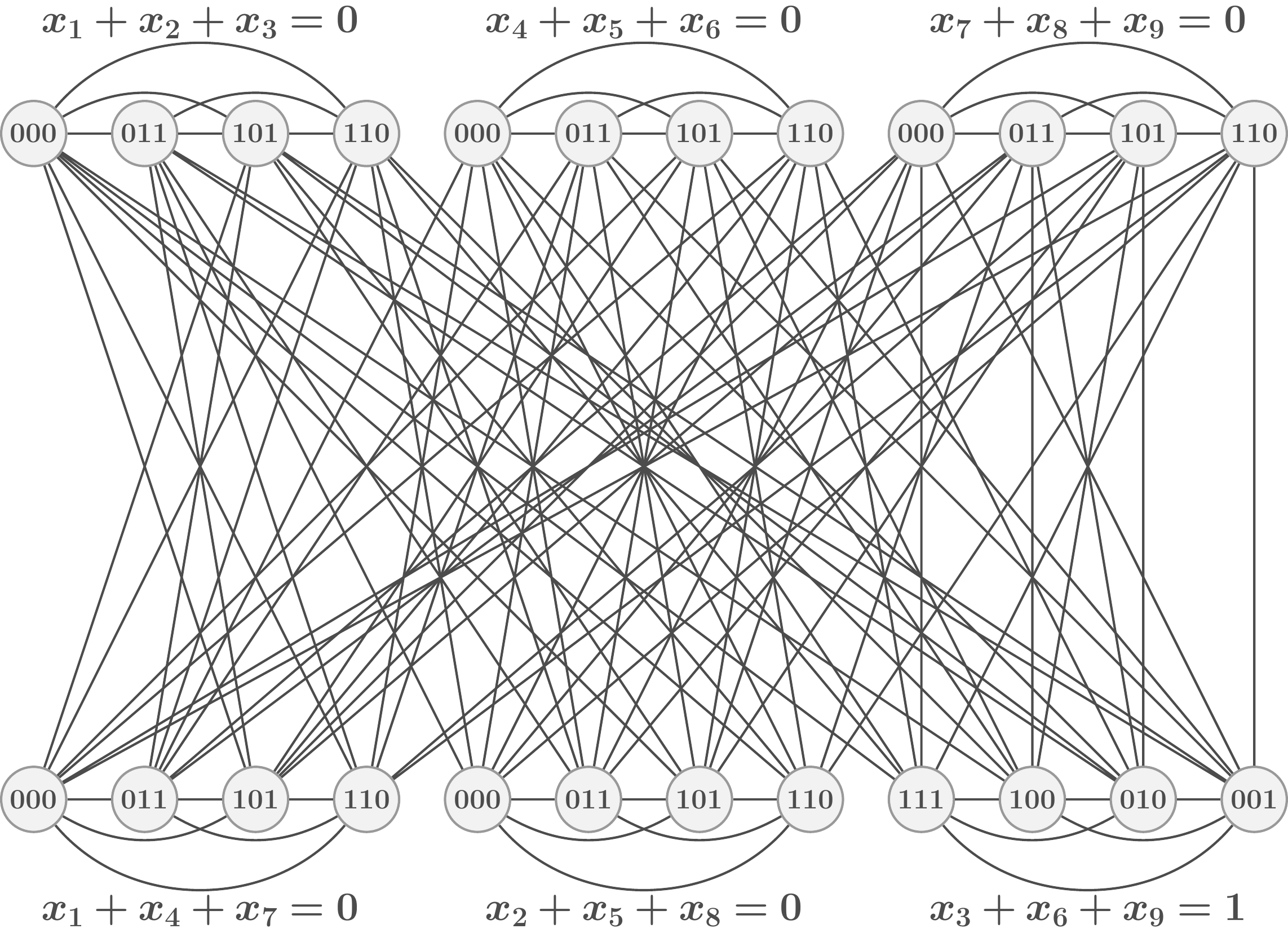}
\caption{$G(\FF)$ for the Mermin magic square game~(\ref{eqn:BCS}).}\label{fig:qiso1}
\end{center}
\end{figure}

\begin{figure}[h!]
\begin{center}
\includegraphics[scale=.52]{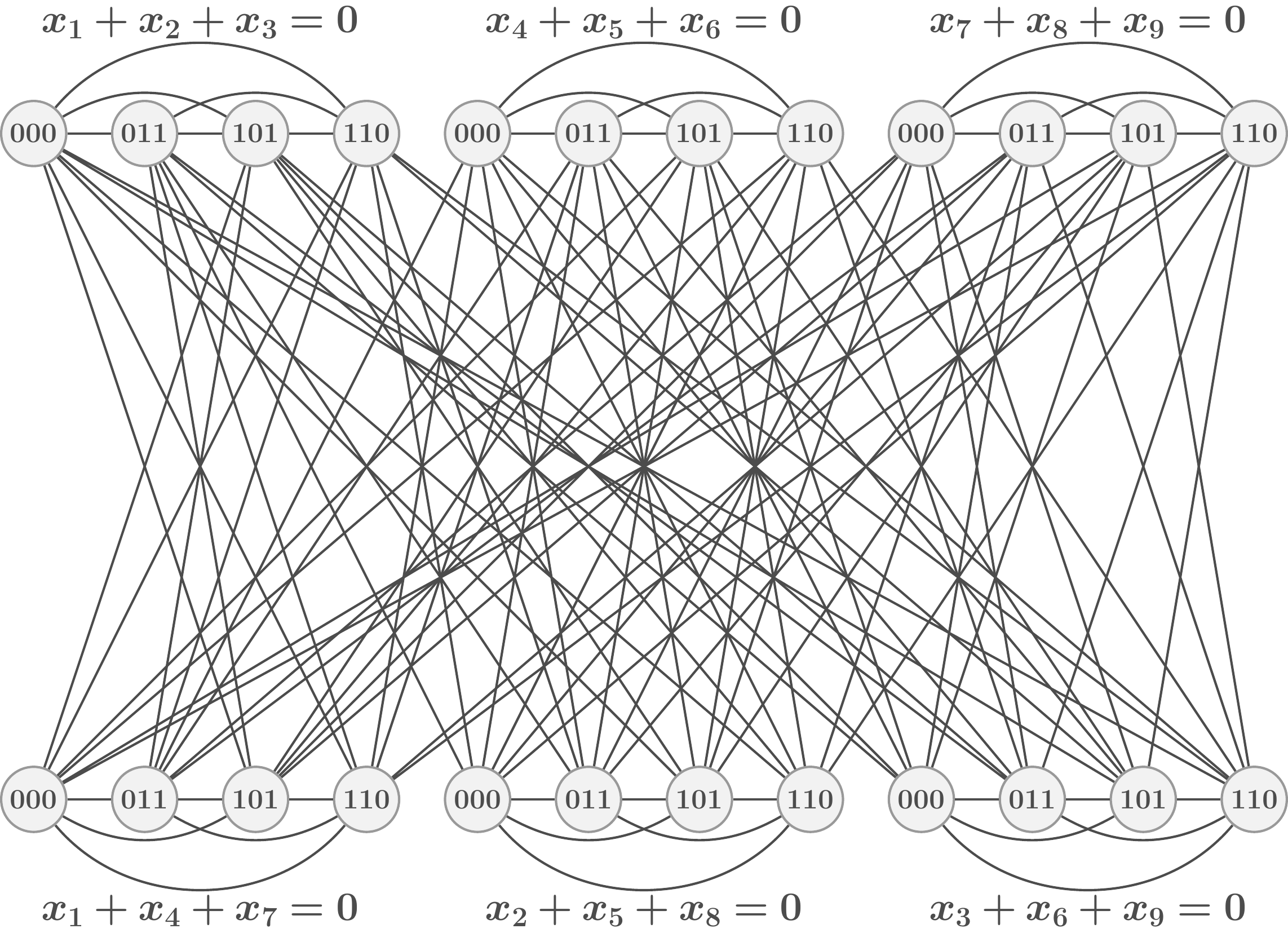}
\caption{$G(\FF_0)$ for the Mermin magic square game~(\ref{eqn:BCS}).}\label{fig:qiso2}
\end{center}
\end{figure}

We note here that the first separating example was slightly different than the one presented above. It was a version of the celebrated CFI construction, named after Cai, F\"urer, and Immerman \cite{CFI}. The original
CFI construction was designed to produce pairs of non-isomorphic graphs
that cannot be distinguished by the $d$-dimensional Weisfeiler-Lehman
algorithm for any fixed $d$. The CFI construction was reinterpreted by
Atserias, Bulatov, and Dawar \cite{ABD} to view it as an encoding of special
systems of linear equations over $\mathbb{Z}_2$, where each
variable appears in precisely two equations. Our first separating
example was literally the CFI construction corresponding to a system of
linear equations as in \cite{ABD}, in which each variable appears in exactly
two equations, and that is classically unsatisfiable over $\mathbb{Z}_2$ but
quantum satisfiable.
The Mermin-Peres
magic square game gives rise to such a system of linear equations.
When applied to the constraint system describing the magic square, this construction produced graphs with several
hundred vertices. The final construction which we described above is a
simplified version of this, in which several vertices have been merged
together, and several others have been removed, without changing the
outcome. The final graphs have a few dozen vertices. As it turns
out, this streamlined version of the construction is quite similar to
the FGLSS reduction from the theory of hardness of approximation \cite{FGLSS},
which interpreted in this context is a reduction from the feasibility
problem for arbitrary systems of linear equations over $\mathbb{Z}_2$ (without any
restriction on the number of occurrences of each variable) to the graph
isomorphism problem. The FGLSS construction was also
used in the context of the graph isomorphism problem in \cite{odonnell}.

\subsection{Separating quantum and quantum commuting isomorphism}

We can also apply the techniques of the previous section to show that the existence of quantum commuting strategies for a linear BCS game can be reduced to the existence of a quantum commuting strategy for a corresponding isomorphism game. In particular, we have the following analog of Theorem~\ref{thm:qBCS2qiso}:

\begin{theorem}\label{thm:qcBCS2qciso}
Let $\FF$ be a linear BCS with $m$ constraints. Then the following are equivalent:
\begin{enumerate}[(i)]
\item The BCS game for $\FF$ has a perfect quantum commuting strategy;
\item The graphs $\GF$ and $\GFO$ are quantum commuting isomorphic;
\item There exists a tracial packing of $\GF$ of value $m$;
\item $\alpha_{qc}(\GF) = m$.
\end{enumerate}
\end{theorem}
\proof
The proof follows the same format as that of Theorem~\ref{thm:qBCS2qiso}. Indeed, note that proof $(i) \Rightarrow (ii)$ from above is independent of the type of strategy used, i.e., it works just as well for the quantum commuting case. The proof of $(ii) \Rightarrow (iii)$ is identical as well, with tracial packing replacing projective packing. To show that $(iii) \Rightarrow (i)$, the main difference is to just use $s(E_{(\ell,f)})$ in place of $\rk(E_{(\ell,f)})/d$, and the fact that for a faithful tracial state $s$, the equality $\sum_{i \in [r]} s(E_i) = 1$ implies that $\sum_{i \in [r]} E_i = I$ for any mutually orthogonal projectors $E_1, \ldots, E_r$. It is then easy to see that the projectors $E_{(\ell,f)}$ satisfy Lemma~\ref{lem:synchstrats} and thus the BCS game for $\FF$ has a perfect quantum commuting strategy. The equivalence of $(iii)$ and $(iv)$ is identical to the above, just using the tracial packing analog of Lemma~\ref{lem:projpackequal}.\qeds

For our main results on quantum commuting isomorphism, we will use the above along with the following two results of Slofstra~\cite{slofstra16}:
\begin{theorem}[Slofstra]\label{thm:slofstra1}
There is a linear BCS game that has a perfect quantum commuting strategy but no perfect quantum strategy.
\end{theorem}
\begin{theorem}[Slofstra]
It is undecidable to determine if a linear BCS game has a perfect quantum commuting strategy.
\end{theorem}
From our Theorem~\ref{thm:qcBCS2qciso} and the above two theorems of Slofstra we immediately obtain the desired corollaries:
\begin{corollary}\label{cor:qcisonotqiso}
There exist graphs $G$ and $H$ such that $G \cong_{qc} H$ but $G \not\cong_q H$.
\end{corollary}
\begin{corollary}\label{cor:cqisoundec}
It is undecidable to determine if two graphs are quantum computing isomorphic.
\end{corollary}

Unfortunately, the linear binary constraint systems that Slofstra uses to prove Theorem~\ref{thm:slofstra1} are too large to produce graphs of reasonable size, and thus we cannot include any specific examples for Corollary~\ref{cor:qcisonotqiso}. Recall that Theorem~\ref{thm:ns1} states that non-signalling isomorphism and fractional isomorphism are equivalent relations, and further recall that the latter is known to be polynomial time decidable. Thus Corollary~\ref{cor:cqisoundec} implies that quantum commuting isomorphism and non-signalling isomorphism are distinct relations:
\begin{corollary}
There exist graphs $G$ and $H$ such that $G \cong_{ns} H$ but $G \not\cong_{qc} H$.
\end{corollary}
We note that using Slofstra's undecidability result to prove that quantum commuting isomorphism and non-signalling isomorphism are not equivalent is overkill in the extreme. In fact, one can show this more directly using results from our upcoming work~\cite{qiso2}

Combining the results of this section with Theorem~\ref{thm:result3}, we see that isomorphism, quantum isomorphism, quantum commuting isomorphism, and non-signalling isomorphism are all distinct relations, i.e., none of the implications in Equation~(\ref{eqn:imps}) can be reversed. It is interesting how much variation there is in the complexity of these four relations. The weakest among them, non-signalling isomorphism, is polynomial time decidable, but the next strongest, quantum commuting isomorphism, is undecidable. Of course isomorphism itself was recently shown by Babai~\cite{B15} to be decidable in quasipolynomial time. Lastly, the complexity of deciding quantum isomorphism remains open.

\subsection*{Acknowledgements}
AA was partially funded by the European Research Council (ERC) under the
European Union's Horizon 2020 research and innovation programme, grant
agreement ERC-2014-CoG 648276 (AUTAR), and by MINECO through
TIN2013-48031-C4-1-P (TASSAT2).
LM is supported by UK EPSRC under grant EP/L021005/1. 
DR is supported by Cambridge Quantum Computing Ltd.~and the Engineering and Physical Sciences Research Council of the United Kingdom (EPSRC), as well as Simone Severini and Fernando Brandao.
RS is partially supported by grant GA \v{C}R P202-12-G061 and by grant LL1201 ERC CZ of the Czech Ministry of Education, Youth and Sports. 
SS is supported by the Royal Society, the EPSRC, and the National Natural Science Foundation of China~(NSFC).
AV is supported in part by the Singapore National Research Foundation under NRF RF Award No. NRF-NRFF2013-13. 
Part of this work was done while AA, DR, and SS were visiting the Simons Institute for the Theory of Computing.

\bibliographystyle{plainurl}

\bibliography{Quantum_Isomorphisms.bbl}

\end{document}